\documentclass[aps,pra,preprint,a4paper,showpacs,amsmath,amssymb]{revtex4-1}

\usepackage[utf8]{inputenc}

\usepackage{pdfsync}

\usepackage{amsmath}
\usepackage{amssymb}
\usepackage{latexsym}
\usepackage{mathrsfs}

\usepackage{boxedminipage}

\usepackage{ifpdf}

\ifpdf
\usepackage[pdftex]{graphicx}
\else
\usepackage{graphicx}
\fi

\ifpdf
\DeclareGraphicsExtensions{.pdf, .jpg, .tif}
\else
\DeclareGraphicsExtensions{.eps, .jpg}
\fi

\usepackage{float}

\begin{document}

\title{Structure of the Alkali-metal-atom-Strontium molecular ions: towards photoassociation and formation of cold molecular ions}

\author{M. Aymar$^{1}$, R. Gu\'erout$^{2}$,  and O. Dulieu$^{1}$}
\affiliation{$^{1}$Laboratoire Aim\'e Cotton, CNRS, UPR3321, B\^at. 505, Univ Paris-Sud, 91405 Orsay Cedex, France\\
$^{2}$Laboratoire Kastler-Brossel, CNRS, ENS, Univ Pierre et Marie Curie case 74,
Campus Jussieu, F-75252 Paris Cedex 05, France}
\email[O. Dulieu: ]{olivier.dulieu@u-psud.fr}

\date{\today}

\begin{abstract}
The potential energy curves, permanent and transition dipole moments, and the static dipolar polarizability, of molecular ions composed of one alkali-metal atom and a Strontium ion are determined with a quantum chemistry approach. The molecular ions are treated as effective two-electron systems and are treated using effective core potentials including core polarization, large gaussian basis sets, and full configuration interaction. In the perspective of upcoming experiments aiming at merging cold atom and cold ion traps, possible paths for radiative charge exchange, photoassociation of a cold Lithium or Rubidium atom and a Strontium ion are discussed, as well as the formation of stable molecular ions.
\end{abstract}

\maketitle




\section{Introduction}

As illustrated by two recent review papers \cite{carr2009,dulieu2009}, the success story of researches on ultracold neutral molecules is developing for more than ten years since their first detection in an ultracold cesium gas \cite{fioretti1998,takekoshi1998,takekoshi1999}. Ultracold molecules indeed offer exciting prospects for fundamental quantum physics and physical chemistry, as well as for interdisciplinary physics and possible applications and for technology. The ultimate control of their external degrees of freedom down to kinetic energies equivalent to temperatures far below 1~mK, and of their internal degrees of freedom at the single (lowest) quantum level \cite{nikolov2000,sage2005,deiglmayr2008a,viteau2008,ni2008,danzl2008,lang2008a,danzl2010,shuman2010}, is a crucial path towards the realization of a degenerate quantum gas of ultracold molecules in their absolute ground state. Ensembles of ultracold molecules in well-defined quantum states allow for the observation of cold collisions between atoms and molecules \cite{zahzam2006,staanum2006,hudson2008,hummon2010}. The control of elementary chemical reactions is now at reach \cite{cvitas2005,quemener2008,krems2008,bell2009,kirste2010,ni2010,ospelkaus2010a}, while new light on the quantum behavior of few-body systems is brought through the demonstration of predicted universal laws \cite{efimov1970,efimov2010}. Many-body physics could be engineered with ultracold polar molecules serving as models for spin lattices with tunable interactions \cite{micheli2006}, or as quantum bits in the perspective of the long-term achievement of a quantum information device \cite{demille2002,rabl2006}. The unprecedented accuracy on transition frequency measurements induced by their ultra-low kinetic energy makes polar molecules suitable for testing fundamental theories, like for instance the variance of the electron-to-proton mass ratio $m_e/m_p$ \cite{demille2008,kajita2008,zelevinsky2008} or electron dipole moment (EDM) \cite{hudson2008,tarbutt2009}.

Besides, molecular ions are well-known for their crucial role in the evolution of many environments like stellar atmospheres or interstellar media \cite{klemperer2006,herbst2008}, and it is highly desirable to create samples of cold molecular ions in order to learn all the details of their quantum dynamics and to control their reactivity \cite{gerlich2008b,willitsch2008}. In this matter, electromagnetic traps for ions have long been used \cite{gerlich2008b}. Novel developments occurred with the demonstration of sympathetic cooling of molecular ions with laser-cooled atomic ions thus forming so-called ion crystals where individual ions are kept at a well-defined positions in space \cite{willitsch2008}. It is nowadays possible to drive such ions into selected rotational level through additional laser-pumping mechanisms \cite{schneider2010,staanum2010,tong2010}. Just like neutral molecules, cold molecular ions are promising candidates for $m_e/m_p$ variance \cite{schiller2005} or EDM \cite{meyer2006,meyer2009} measurement.

For such studies, the prerequisite step is obviously the formation of a cold and trapped sample of the desired species of molecular ions which would be sympathetically cooled by laser-cooled and trapped atomic ions. It is now established that excited ions in the trap and surrounding molecular species introduced on purpose easily react together to produce molecular ions, as demonstrated for the first time in Ref.\cite{molhave2000} with Mg$^+$ and H$_2$ yielding cold and trapped MgH$^+$ ions. With the developing experimental trend of merging traps of laser-cooled atoms and ions like Yb-Yb$^+$ \cite{grier2009}, Rb-Yb$^+$ \cite{zipkes2010,zipkes2010a}, Rb-Ba$^+$ \cite{schmid2010}, new opportunities could be investigated to create diatomic molecular ions \textit{in situ} via \textit{e.g.} the photoassociation-stabilization sequence $A+B^+ + h\nu \rightarrow (AB^+)^* \rightarrow AB^+ + h\nu'$, or via magnetoassociation involving Feshbach resonances \cite{idziaszek2009}, just like routinely done with cold atoms \cite{jones2006,dulieu2009,chin2010}. Charge exchange between the neutral and ionic atomic species should also be considered, as explored in the Na-Ca$^+$ system \cite{makarov2003,idziaszek2009}.

The main drawback for accurate theoretical modeling of these formation mechanisms often arises from the poor knowledge of the molecular structure of such mixed species beyond their electronic ground state. As mentioned in our previous work \cite{guerout2010} (hereafter referred to as paper I), this is particularly true for the diatomic systems, either neutral or charged, formed from an alkali-metal atom and Strontium. The Strontium ion is indeed an attractive system for trapping and laser-cooling  experiments (see for instance Refs.\cite{removille2009,removille2009a}). Following paper I, the present work is devoted to the theoretical determination of the electronic structure of ($A$-$^{87}$Sr)$^+$ ions (where $A$= $^{6}$Li, $^{23}$Na, $^{39}$K, $^{85}$Rb, $^{133}$Cs) using the method briefly recalled in Section \ref{sec:method}. Results for potential energy curves and permanent and transition dipole moments are given in Section \ref{sec:pot} and \ref{sec:other} respectively. In Section \ref{sec:discussion} we explore the possibility to create cold ($A$-Sr)$^+$ ions via photoassociation followed by spontaneous decay, and the expected reactivity of these ions with surrounding cold neutral atoms. Atomic units for distances (1~a.u.$= a0 =$ 0.052 917 720 859~nm), energies (1~a.u.$=2R_{\infty}=
219 474.631 370 54$~cm$^{-1}$), and dipole moment (1~a.u. =2.541 58 Debye) will be used throughout the paper, except otherwise stated.

\section{Computational approach}
\label{sec:method}

The present calculations actually represent an intermediate step of those performed in paper I for neutral $A$-Sr molecules, which were modeled as effective three-electron systems. The $A$-Sr$^+$ ion has two valence electrons, and its electronic structure is, at least in terms of values of quantum numbers, similar to the one of alkali dimers. Therefore we describe the $A$-Sr$^+$ ion as an effective two-electron system following the same steps than in paper I and in Ref.\cite{aymar2005}, which are briefly recalled below.

Each of the closed-shell ionic cores $A^+$ and Sr$^{2+}$ are represented by an effective core potential (ECP) from Durand and Barthelat \cite{durand1974,durand1975} and Fuentealba \textit{et al.} \cite{fuentealba1985,fuentealba1987} respectively, completed by the core-polarization potential (CPP) depending on the angular momentum $\ell$ of the valence electron, along the lines of ref.\cite{muller1984} and
revisited by Gu\'erout \textit{et al.} \cite{guerout2010a}. The valence electron wave functions are expanded on a Gaussian orbitals centered at each ionic core. The molecular  orbitals are determined   by restricted
Hartree-Fock  single electron calculations, including the CPP term yielding the potential curves for the relevant molecular double-charged cations. Then a full configuration interaction (FCI) is performed for each relevant molecular symmetry, providing potential curves, and permanent and transition dipole moments. The calculations are performed through an automatic procedure \cite{aymar2005} based on the  the CIPSI package (Configuration Interaction by Perturbation of a multiconfiguration wave function Selected Iteratively) \cite{huron1973}. As mentioned in paper I we use for the Sr$^+$ ion a large uncontracted basis set \{$7s6p7d$\} derived from the \{$5s5p6d1f$\} used in Ref. \cite{boutassetta1996}, and \{$8s7p4d$\}, \{$7s6p5d$\}, \{$7s5p7d$\}, \{$7s4p5d$\}, and \{$7s4p5d$\} uncontracted basis sets for Li, Na, K, Rb, and Cs, respectively. All exponents of the Gaussian orbitals, as well as the cut-off radii and the core polarizabilities \cite{coker1976,wilson1970} appearing in the CPP term for the alkali species, were already reported in paper I and are not repeated here. The cut-off radii for the Sr$^{2+}$ CPP related to the \{$7s6p7d$\} basis set are slightly adjusted compared to paper I, \textit{i.e.} $\rho_s=2.13035$, $\rho_s=2.183$, $\rho_s=1.70616$, in atomic units. The quality of the basis sets is settled through several criteria. First of all, the cut-off radii are adjusted to reproduce the experimental energy of the lowest levels of the alkali atoms and and of the Sr$^+$ ion (see Table 3 in paper I). Next, as the lowest dissociation limit of the $A$-Sr$^+$ ions is $A^+$+Sr, the lowest energy level of the -effective two-electron- Sr atom calculated by the FCI above should be well reproduced. As shown in paper I, the present basis set is quite efficient in this respect, as the Sr ground state and $5s5p^3P$, and $5s5p^1P$ excited state energies are calculated with a difference of 185~cm$^{-1}$, 269~cm$^{-1}$, and 434~cm$^{-1}$ compared to the experimental values, respectively. The corresponding excitation energies of the $5s5p^3P$ and $5s5p^1P$ levels are obtained with a difference of 87~cm$^{-1}$, and 262~cm$^{-1}$ compared to the experimental values, which represents an accuracy comparable to the one achieved in the recent high-accuracy calculations on Calcium by Bouissou \textit{et al.} \cite{bouissou2010}.

In addition, the quality of the electronic wave functions of the Sr$^+$ ion can be checked against the calculated static dipole polarizability, which has been studied theoretically for several energy levels in various contexts: the broadening of spectral lines due to collisions with hydrogen atoms \cite{barklem2002}, the long-range interaction between the Sr$^+$ ion and various atomic species \cite{patil1997,mitroy2008a}, or the search for stable and reproducible atomic frequency standards based on the  $5s-4d_{5/2}$ clock transition in $^{88}$Sr$^+$ \cite{margolis2003,madej2004,jiang2009}. All these calculations are based on the summation over oscillator strengths at various levels of approximations, namely including transitions towards the continuum \cite{patil1997,mitroy2008a} or not \cite{barklem2002,madej2004,jiang2009}. Here we use the finite field method \cite{cohen1965} for the contribution of the valence electron, as in Ref.\cite{sadlej1991} and in our previous work on alkali atoms \cite{deiglmayr2008}. More precisely we calculated the energies of  the $2\ell +1$ components of an atomic state with angular momentum $\ell$ in the presence of an electric field varying from 0.002 to 0.004~a.u., and extracted the polarizability from the quadratic dependence of the averaged values  against the electric field magnitude. Finally we add the contribution of the Sr$^{2+}$ core (5.67~a.u after Ref. \cite{coker1976}), as pointed out in Ref.\cite{sternheimer1969}. Our results are reported in Table \ref{tab:pola_sr+} for the $5s$, $4d$, $6s$,  $5d$, and $5p$ levels. Our values for the $5s$ and $4d$ states agree well with all other values except the oldest one of Ref.\cite{sadlej1991}, and are in reasonable agreement with the most recent ones of Ref.\cite{mitroy2008a} for the excited $p$ and $s$ levels. The underestimation visible in our calculations for the $d$ levels could be due to the  absence of $f$ orbitals in our  basis sets.

We calculated with the same approach the polarizability of Sr in its ground state (Table \ref{tab:pola_sr+}) which agrees with less than a 10\% difference with the experimental value of Ref.\cite{schwartz1974}, as well as with the recent theoretical value of Ref. \cite{derevianko2010}. All other theoretical determinations are larger by 20\% or more. A new experimental determination of electric dipole polarizability of strontium then seems desirable to discriminate among all determinations.

\begin{table}[t]
\center
\begin{tabular} {|c|c|c|c|c|c|c|c|}
State&This work& \cite{mitroy2008a}& \cite{madej2004}&\cite{jiang2009}&\cite{patil1997}& \cite{barklem2002}&
\cite{sadlej1991}\\ \hline
$5s$&92.7 &89.8  &84.6&91.3&91.47&93.1&130.5\\
$4d$&53.3 &61.77 &53.7&62  &     &57  &   \\
$5p$&-49.2&-23.13&    &    &     &-32 &\\
$6s$&1305 &1089  &    &    &     &    &  \\
$5d$&1294 &2099  &    &    &     &    &\\
$6p$&-2016&-2056 &    &    &     &    &\\ \hline \hline
&This work&\cite{schwartz1974}&\cite{desclaux1981}-DF-HF&\cite{fricke1986}& \cite{thorhallsson1968}&\cite{sadlej1991}&\cite{derevianko2010} \\
$5s^2$&200&186$\pm 14$&228-241&186&225&245.7&197.2 \\ \hline
\end{tabular}
\caption{Static dipole polarizability of $5s$ $4d$, $5d$ and $6p$ states of  Sr$^+$ ion (in atomic units)  compared with other calculated values. We added the Sr$^{2+}$ polarizability of 5.67~a.u. \cite{coker1976} to the values of Refs.\cite{madej2004,barklem2002}. The indexes DF and HF hold for Dirac-Fock and Hartree-Fock level of calculation, respectively. The static dipole polarizability for the Strontium ground state is also displayed, compared to several theoretical values, and an experimental result \cite{schwartz1974}.}
\label{tab:pola_sr+}
\end{table}

\section{Potential curve calculations}
\label{sec:pot}

\begin{figure}[t]
  \begin{center}
    \includegraphics[width=0.6\textwidth]{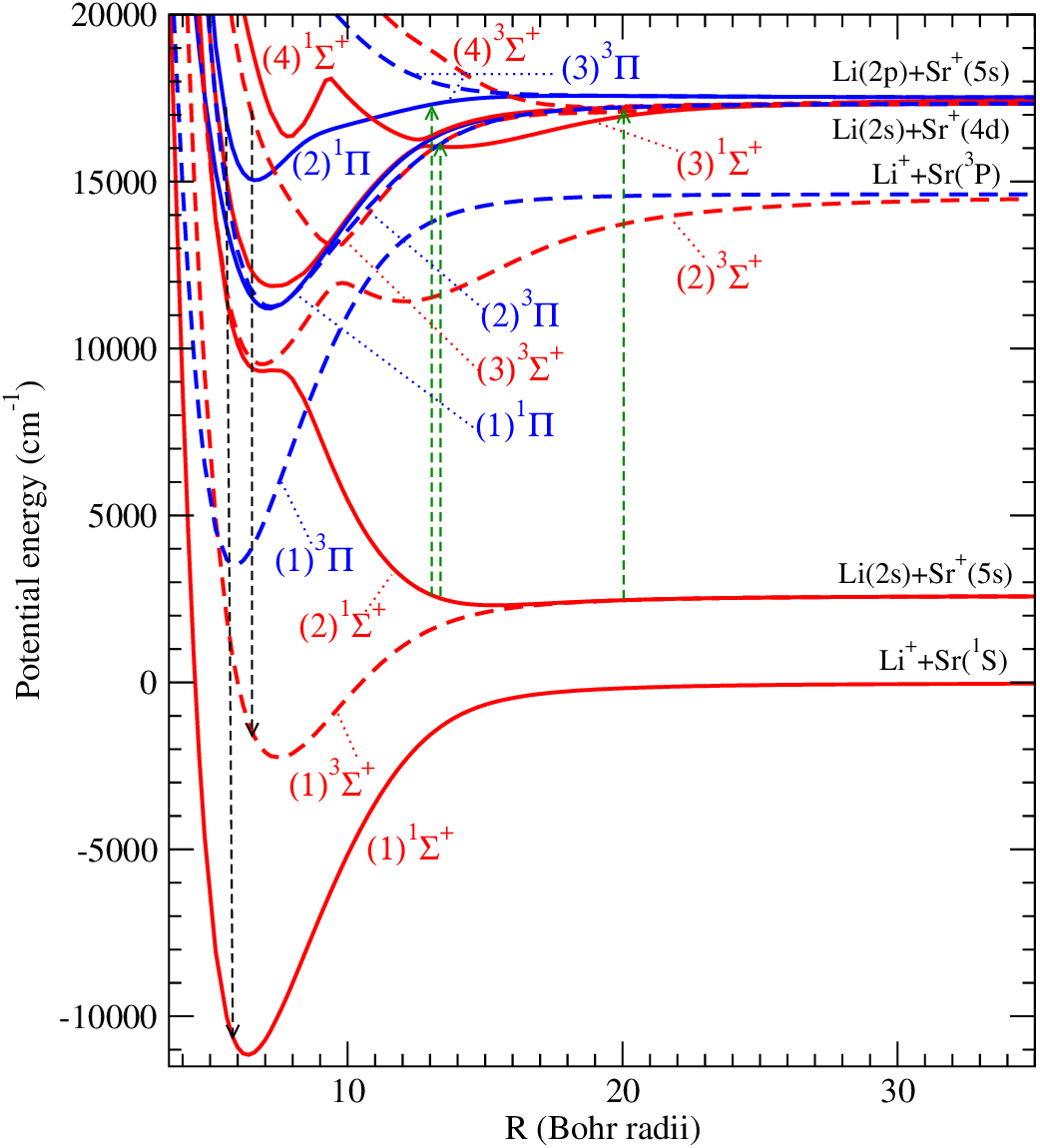}
  \end{center}
  \caption{Potential energy curves of the LiSr$^+$ molecule computed in the present work. $^1\Sigma^+$: full red lines; $^3\Sigma^+$: dashed red lines; $^1\Pi$: full blue lines; $^3\Pi$: dashed blue lines. Vertical arrows suggest possible photoassociation (upwards) of cold Li and Sr$^+$ particles and spontaneous decay paths (downwards) to create cold LiSr$^+$ molecules. Note that curves of $\Delta$ symmetry are not drawn for clarity sake.}
  \label{fig:Li}
\end{figure}

\begin{figure}[t]
  \begin{center}
    \includegraphics[width=0.6\textwidth]{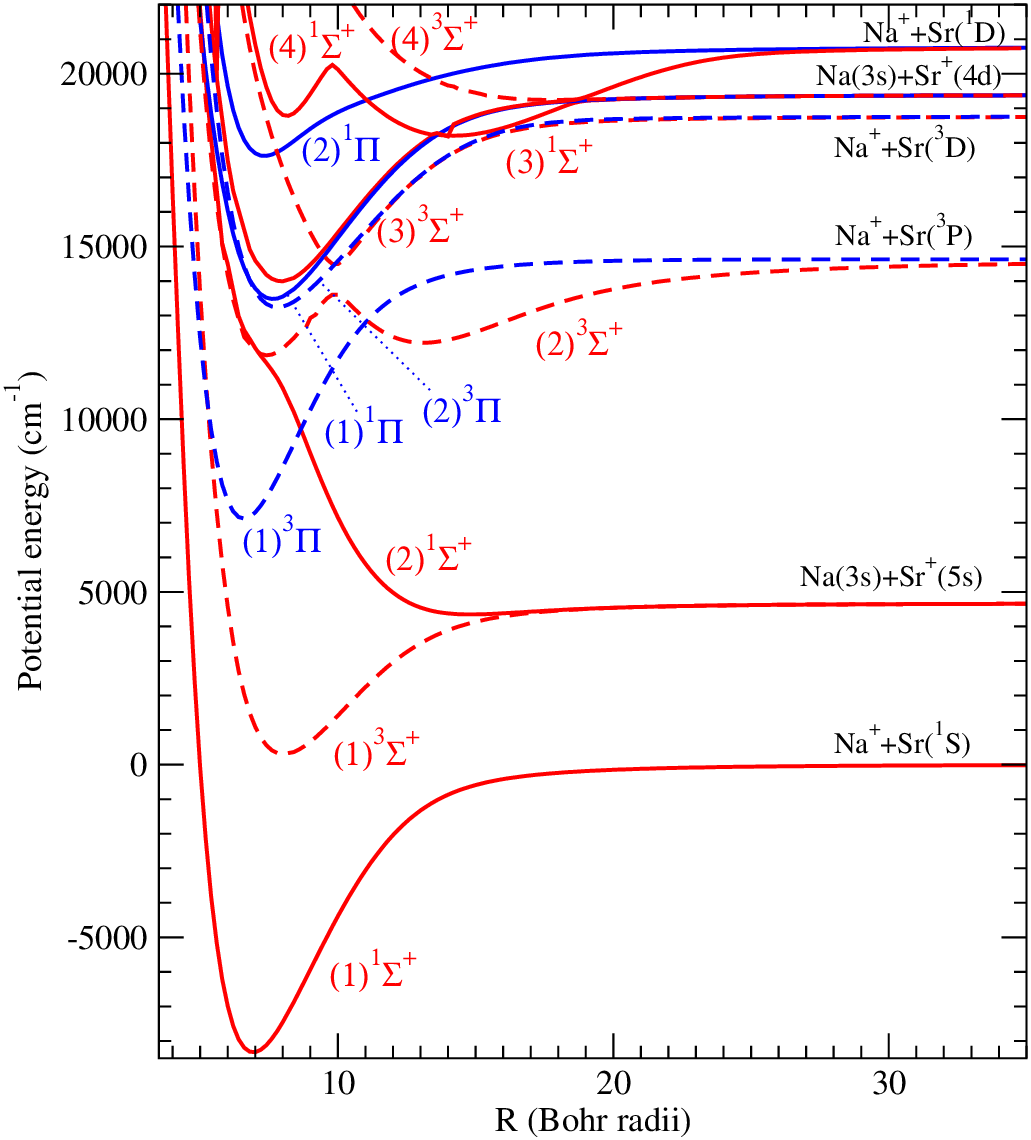}
  \end{center}
  \caption{Potential energy curves of the NaSr$^+$ molecule computed in the present work. Line types as in Fig.\ref{fig:Li}. Note that curves of $\Delta$ symmetry are not drawn for clarity sake.}
  \label{fig:Na}
\end{figure}

\begin{figure}[t]
  \begin{center}
    \includegraphics[width=0.6\textwidth]{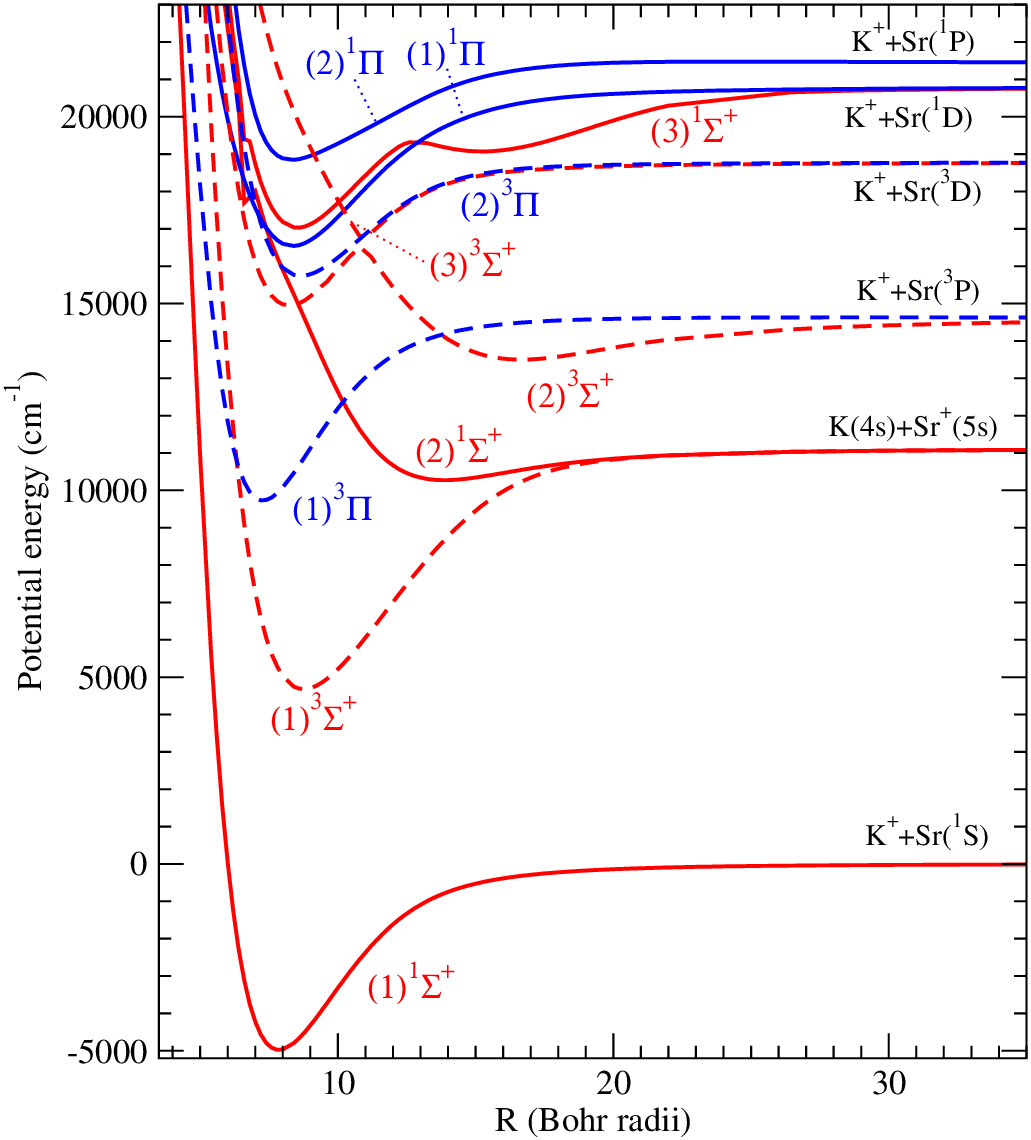}
  \end{center}
  \caption{Same as Fig. \ref{fig:Na} for KSr$^+$.}
  \label{fig:K}
\end{figure}

\begin{figure}[t]
  \begin{center}
    \includegraphics[width=0.6\textwidth]{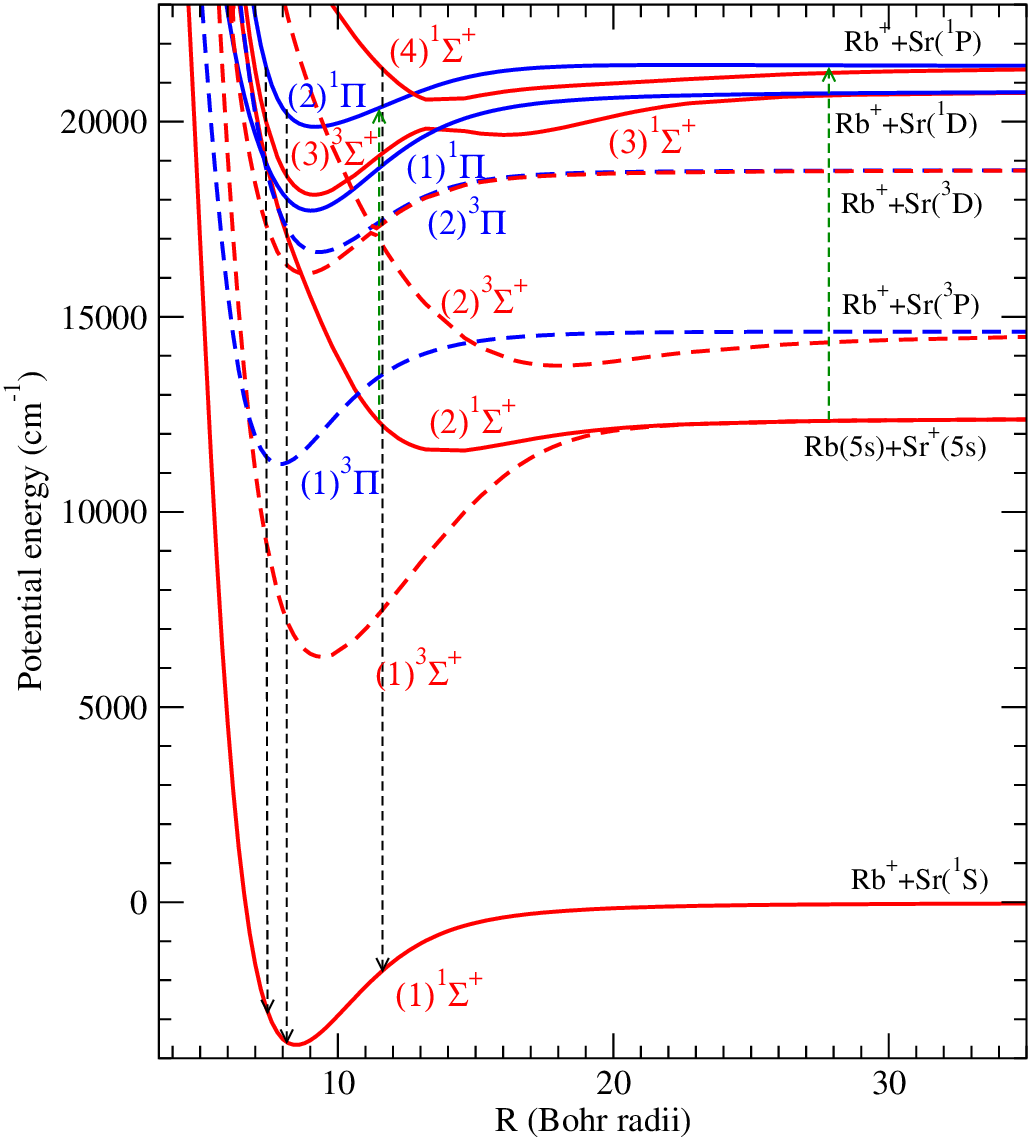}
  \end{center}
  \caption{Same as Fig.\ref{fig:Li} for RbSr$^+$.}
  \label{fig:Rb}
\end{figure}

\begin{figure}[t]
  \begin{center}
    \includegraphics[width=0.6\textwidth]{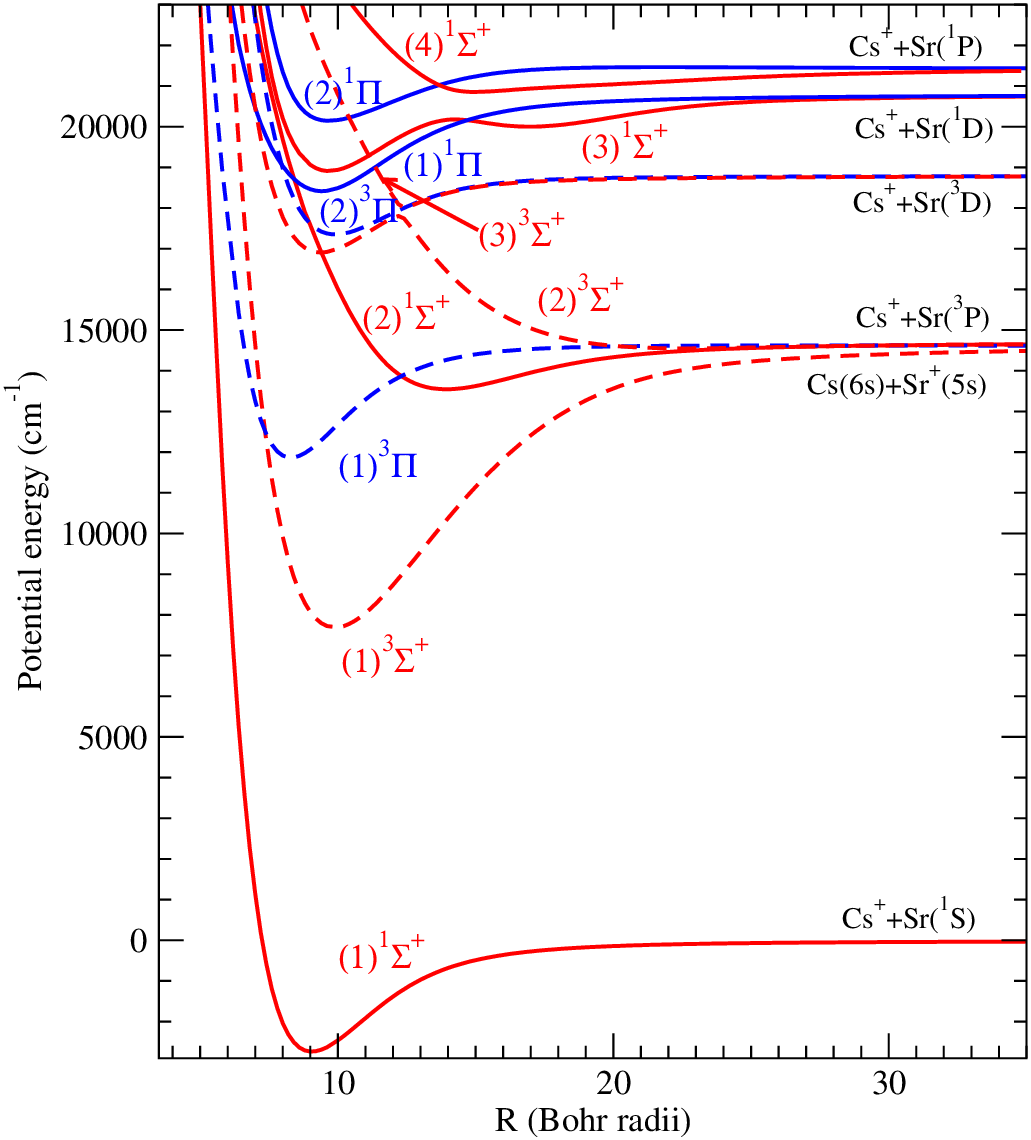}
  \end{center}
  \caption{Same as Fig. \ref{fig:Na} for CsSr$^+$.}
  \label{fig:Cs}
\end{figure}

An overview of the results obtained from the procedure above for the potential curves of the $A$-Sr$^+$ ions is provided in Figs. \ref{fig:Li} to \ref{fig:Cs}, while the main spectroscopic constants are reported in Tables \ref{tab:constants_Li} to \ref{tab:constants_Cs}. In the tables, we recall for convenience the dissociation energies of the asymptotic limits as displayed in paper I. Let us note that no other results for these systems have been previously published. The corresponding numerical data are provided as additional material accompanying the paper.

\begin{figure}[htb]
  \begin{center}
    \includegraphics[width=0.6\textwidth]{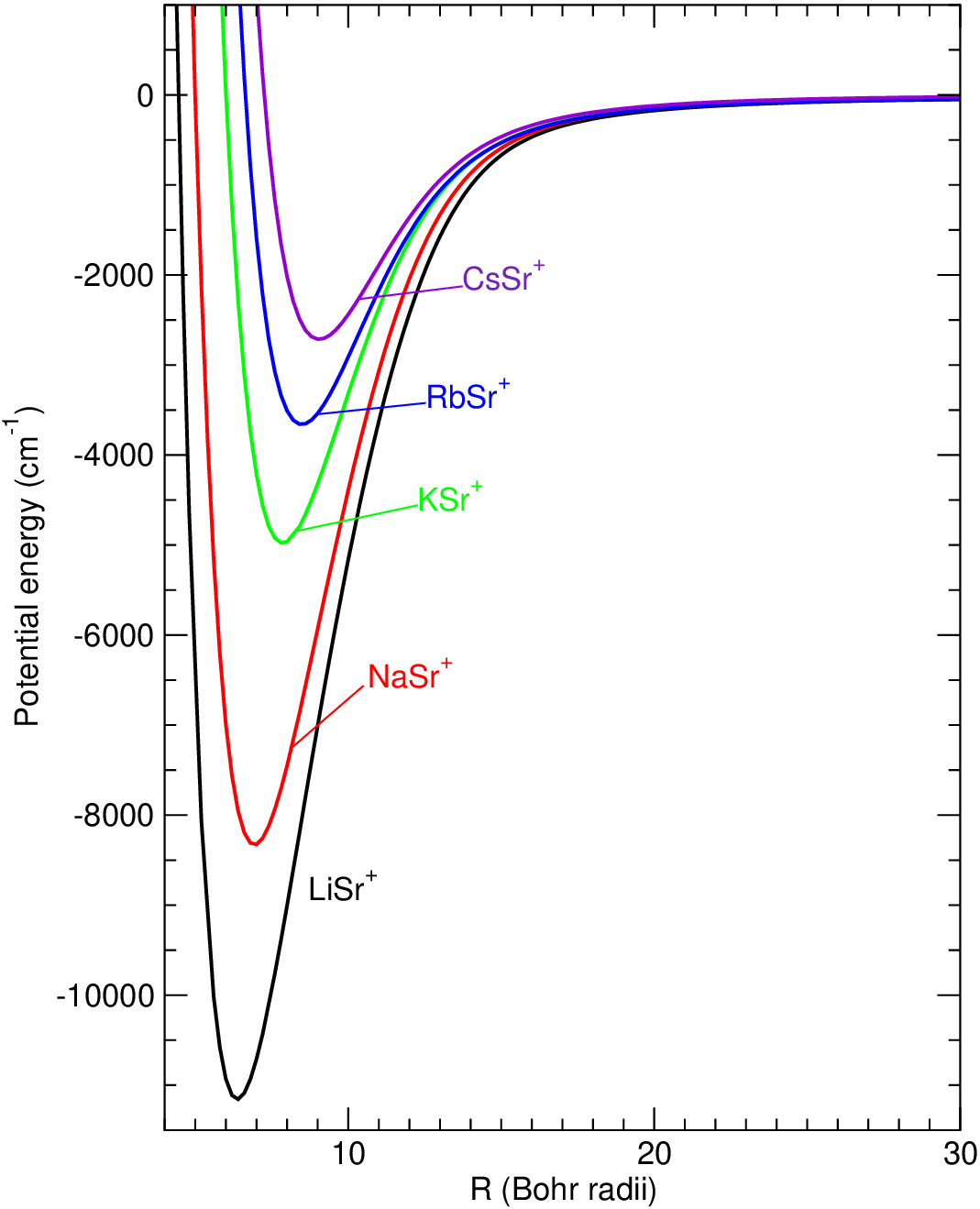}
  \end{center}
  \caption{Potential energy curves of the X$^1\Sigma^+$ ground state of $A$-Sr$^+$ ions, all referring to the energy origin fixed at the $A$+Sr$^+$ dissociation.}
  \label{fig:Xall}
\end{figure}

For all ions, the ground state X$^1\Sigma^+$ of the system correlates at large distances to a ground-state neutral Sr atom and a closed-shell alkali-metal ion. The ordering of the well depths is the same than with neutral $A$-Sr molecules \cite{guerout2010}: the deepest wells are by far the LiSr$^+$ and NaSr$^+$ ones (11157~cm$^{-1}$, and 8330~cm$^{-1}$, respectively), while the well depths of KSr$^+$, RbSr$^+$, CsSr$^+$ are significantly smaller (Fig. \ref{fig:Xall}). This can be understood as the energy spacing between the two lowest dissociation limits increases with increasing alkali-metal atom mass: the X$^1\Sigma^+$ state is less and less repelled by the next (2)$^1\Sigma^+$ state, and the $X$ state behaves more and more like a van der Waals molecule composed of two closed-shell systems $A^+$ and Sr($5s^2$). The equilibrium distances are comparable to the typical values of alkali dimers (see for instance Ref.\cite{stwalley2010} where a sketch of potential curves for all heteronuclear alkali-metal dimers is displayed) as the molecules are formed by two open-shell partners.
At the second dissociation limit, one valence electron is attached to each atomic partner, giving rise to a singlet and a triplet state at short internuclear distances. Despite this similarity with the lowest electronic states of ground state alkali-metal pairs, the situation is here strikingly different: the lowest triplet state (1)$^3\Sigma^+$ (also referred to as the a$^3\Sigma^+$ state) potential curve is strongly attractive while the accompanying singlet state (2)$^1\Sigma^+$ (also referred to as the A$^1\Sigma^+$ state) potential curve is mostly repulsive, due to the strong repulsion induced by the X$^1\Sigma^+$ ground state. The (2)$^1\Sigma^+$ would actually represent a strong dissociation channel induced by spin-orbit coupling for the (1)$^3\Pi$ state if collisions between alkali-metal ions and neutral Strontium were studied. However, as a reciprocal effect of the larger and larger energy spacing between dissociation limits invoked above, the heavier $A$-Sr$^+$ molecule, the deeper long-range potential well in the (2)$^1\Sigma^+$ potential curve.

The (1)$^3\Sigma^+$ is significantly deeper (several thousands of cm$^{-1}$) than the one of alkali-metal dimers (generally having a few hundreds of cm$^{-1}$ depth, see Ref.\cite{stwalley2010} again). Indeed, the charge-induced dipole long-range attraction (varying as $R^{-4}$, where $R$ is the internuclear distance) between Sr$^+$ and the alkali-metal atom is stronger than in the alkali-metal pairs (varying as $R^{-6}$), and dominates the (repulsive) exchange interaction down to short distances. As the static dipole polarizability (determining the strength of the long-range $A$-Sr$^+$ interaction) increases with increasing alkali-metal atom mass, the depth of the (1)$^3\Sigma^+$ increases, and becomes even larger than the X$^1\Sigma^+$ well depth for KSr$^+$, RbSr$^+$, CsSr$^+$. Note that this trend is similar to what is also observed for the RbBa$^+$ ion \cite{krych2010}.

The combination of the excitation energies of Sr$^+$, of Sr, and of the alkali-metal atoms induces the main differences for excited molecular states among the five different $A$-Sr$^+$ species. For instance, the LiSr$^+$ molecule is the only one where the dissociation limit with the alkali-metal atom excited in its lowest $p$ level has an energy larger than the one involving Sr($5s5p\,^1P$). The Li($2p$)+Sr$^+(5s)$ dissociation limit actually lies only 178~cm$^{-1}$ above the Li($2s$)+Sr$^+(4d)$ one. Similarly, the LiSr$^+$ and NaSr$^+$ species are the only ones in which the dissociation limit involving the Sr$^+(4d)$ level has a binding energy smaller than the one involving Sr($5s5p\,^1P$). In this respect, the KSr$^+$, RbSr$^+$, CsSr$^+$ exhibit similar patterns in their excited states. Let us note that due to these differences, the lettered labels A, B, C, D, E, a, b, c of the molecular states, which is historically related to their energy ordering, will not always be attached to the same dissociation limits in the $A$-Sr$^+$ series. The consequences of such characteristics for the dynamics of the $A$-Sr$^+$ systems will be further addressed in Section \ref{sec:discussion} below.

Among all the excited states, there are two cases where the corresponding potential curves exhibit remarkable features. First the $(2)^3\Sigma^+$ curve is strongly repelled by the attractive $(1)^3\Sigma^+$ one so that it strongly interact through an avoided crossing with the $(3)^3\Sigma^+$ curve located in the 9-12~a.u. distance range. For all species but LiSr$^+$, it induces a strong mixing of the $^3P$ and $^3D$ states of Strontium, starting from large distances. This will undoubtedly lead to significant predissociation of the $(3)^3\Sigma^+$ bound levels, without charge exchange. This creates a second long-range well in the $(2)^3\Sigma^+$ curve, which almost disappears in CsSr$^+$ due to the continuous decrease of the ionization potential of the alkali species with increasing mass. In the LiSr$^+$, this would actually represent a channel for charge exchange, as the $(3)^3\Sigma^+$ is correlated to the Li($2s$)+Sr$^+$($4d$) limit.

Second, the pair of states $(3)^1\Sigma^+$ and $(4)^1\Sigma^+$ show a similar pattern while slightly more complicated. For the KSr$^+$, RbSr$^+$, and CsSr$^+$ species, these two states dissociate toward the neighboring $A^+$+Sr($^1D$) and $A^+$+Sr($^1P$) limits, resulting into an avoided crossing in the 12-15~a.u. distance range, with a second long-range potential well in the $(3)^1\Sigma^+$ curve. In LiSr$^+$ a similar pattern is predicted between these two states, except that their interaction corresponds to an exchange of excitation for the Li atom, due to the different ordering of the excitation energies in the various species. In between, the NaSr$^+$ ion should exhibit a curious feature: the interaction between these two states results from the proximity of the Na($3s$)+Sr$^+$($4d$) and Na$^+$+Sr($^1D$) limits, so that two successive avoided crossings are visible. Our results suggest that these crossings are quite small, so that it could well be that these two states behave independently in a collision process. The $(4)^1\Sigma^+$ is expected to have a double-well structure as well, which disappears in RbSr$^+$ and CsSr$^+$. These features are actually at the limit of the capabilities of our computations and would require a much larger basis set for an appropriate representation.
\begin{table}[t]
\center
\begin{tabular} {|c|c|c|c|c|c|c|c|} \hline
State          &Label& Diss. limit          &$E_{diss}$(a.u.)&$\delta$(cm$^{-1}$))&$R_e$&$D_e$& $\omega_e$\\ \hline
(1)$^1\Sigma^+$&X    &Li$^+$+Sr($5s^2\,^1S$)&-0.61548        &-184                & 6.37&11126&246\\
(2)$^1\Sigma^+$&A    &Li(2s)+Sr$^+(5s)$     &-0.60350        &0                   &15.33&  277&35.4\\
(1)$^3\Sigma^+$&a    &         "            &     "          &"                   & 7.48& 4865&163\\
(1)$^3\Pi$     &b    &Li$^+$+Sr($5s5p\,^3P$)&-0.54890        &-274                & 5.90&11087&268\\
(2)$^3\Sigma^+$($^*$)&c    &         "            &     "          &"                   & 6.90& \textit{-5086}&190\\
  2$^{nd}$ well        &     &         "            &     "          &"                   &12.1& \textit{-3205}&76\\
   barrier           &     &         "            &     "          &"                   & 9.80& \textit{-2652}& -\\
(1)$^1\Pi$     &B    &Li(2s)+Sr$^+(4d)$     &-0.53641        &0                   & 7.14& 6150&175\\
(3)$^1\Sigma^+$($^*$)&C    &         "            &     "          &"                   & 7.32&\textit{-5479}&130\\
  2$^{nd}$ well        &     &         "            &     "          &"                   & 13.8&\textit{-1297}& -\\
   barrier           &     &         "            &     "          &"                   & 12.9&\textit{-1259}&  \\
(3)$^3\Sigma^+$&     &         "            &     "          &"                   & 9.54& 5722& 69\\
(2)$^3\Pi$     &     &         "            &     "          &"                   & 7.21& 6098& 58\\
(1)$^1\Delta$  &     &         "            &     "          &"                   & 6.92& 7851&194\\
(1)$^3\Delta$  &     &         "            &     "          &"                   & 6.97& 8003&193\\
(2)$^1\Pi$     &D    &Li(2p)+Sr$^+(5s)$     &-0.53559        &0                   & 6.64& 2460&198\\
(4)$^1\Sigma^+$($^*$)&E    &         "            &     "          &"                   & 7.87&\textit{-1183}&260\\
  2$^{nd}$ well        &     &         "            &     "          &"                   &12.59&\textit{-1255}&157\\
   barrier           &     &         "            &     "          &"                   & 9.32&\textit{+591}&- \\ \hline
\end{tabular}
\caption{Main spectroscopic constants of the lowest electronic states (with their conventional spectroscopic label) of the LiSr$^+$ molecule: the equilibrium distance $R_e$ (in atomic units), the well depth $D_e$ (in cm$^{-1}$) and the harmonic constant $\omega_e$ (in cm$^{-1}$). For convenience, we recall the calculated dissociation energies $E_{diss}$, and their difference with the experimental ones \cite{guerout2010}. The ($^*$) sign denotes double-well potentials, for which the barrier height and the energy at the minimum of each well are displayed (in italic) with respect to the dissociation limit, besides their position in $R$. }
\label{tab:constants_Li}
\end{table}
\begin{table}[t]
\center
\begin{tabular} {|c|c|c|c|c|c|c|c|} \hline
State          &Label& Diss. limit          &$E_{diss}$(a.u.)&$\delta$(cm$^{-1}$))&$R_e$&$D_e$& $\omega_e$\\ \hline
(1)$^1\Sigma^+$&X    &Na$^+$+Sr($5s^2$)     &-0.61548        &-184                &6.92&8330&121\\
(2)$^1\Sigma^+$&A    &Na(3s)+Sr$^+(5s)$     &-0.594215       &0                   &14.8& 319&21\\
(1)$^3\Sigma^+$&a    &         "            &     "          &"                   &8.03&4363&83\\
(1)$^3\Pi$     &b    &Na$^+$+Sr($5s5p\,^3P$)&-0.54890        &-274                &6.58&7491&127\\
(2)$^3\Sigma^+$($^*$)&c    &         "            &     "          &"                   &7.43&\textit{-2765}&98\\
  2$^{nd}$ well        &     &         "            &     "          &"                   &13.1&\textit{-2417}&39\\
   barrier           &     &         "            &     "          &"                   & 9.8&\textit{-1017}&-\\
(2)$^3\Pi$     &     &Na$^+$+Sr($5s4d\,^3D$)&-0.52991        &343                 &7.80&5542&85\\
(1)$^3\Delta$  &     &         "            &     "          &"                   &7.45&7091&98\\
(3)$^3\Sigma^+$&     &         "            &     "          &"                   &9.90&4297&148 \\
(1)$^1\Pi$     &B    &Na(3s)+Sr$^+(4d)$     &-0.52712        &0                   &7.64&5911&90\\
(3)$^1\Sigma^+$($^*$)&C    &         "            &     "          &"                   &7.95&\textit{-5402}&86\\
  2$^{nd}$ well        &     &         "            &     "          &"                   &13.4&\textit{-2536}&- \\
   barrier           &     &         "            &     "          &"                   &12.2&\textit{-2113}&- \\
(1)$^1\Delta$  &     &         "            &     "          &"                   &7.41&7438&98\\
(3)$^3\Pi$     &     &         "            &     "          &"                   &17.8&149& 131\\
(4)$^3\Sigma^+$&     &         "            &     "          &"                   &19.0&175& 91\\
(2)$^3\Delta$  &     &         "            &     "          &"                   &17.9&157& 134\\
(2)$^1\Pi$     &D    &Na$^+$+Sr($5s4d\,^1D$)&-0.52073        &460                 &7.32&3166&90\\
(4)$^1\Sigma^+$($^*$)&E    &         "            &     "          &"                   &8.15&\textit{-2008}&129\\
2$^{nd}$ well        &     &         "            &     "          &"                   &14.3&\textit{-1191}&106\\
1$^{st}$ barrier    &     &         "            &     "          &"                   &9.79 &\textit{-516}&- \\
2$^{nd}$barrier     &     &         "            &     "          &"                   &13.6 &\textit{-1158}&- \\  \hline
\end{tabular}
\caption{Same as Table \ref{tab:constants_Li} for NaSr$^+$. For the (3)$^1\Sigma^+$ state, the barrier actually refers to the position of its pseudocrossing point with the (4)$^1\Sigma^+$ curve. Due to the peculiar shape of the latter potential curve, two barriers are identified.}
\label{tab:constants_Na}
\end{table}
\begin{table}[t]
\center
\begin{tabular} {|c|c|c|c|c|c|c|c|} \hline
State          &Label& Diss. limit          &$E_{diss}$(a.u.)&$\delta$(cm$^{-1}$))&$R_e$&$D_e$& $\omega_e$\\ \hline
(1)$^1\Sigma^+$&X    &K$^+$+Sr($5s^2$)     &-0.61548        &-184                &7.86&4977&81\\
(2)$^1\Sigma^+$&A    &K($4s$)+Sr$^+(5s)$   &-0.56487        &0                   &13.8& 834&26\\
(1)$^3\Sigma^+$&a    &          "          &     "          &"                   &8.75&6427&65\\
(1)$^3\Pi$     &b    &K$^+$+Sr($5s5p\,^3P$)&-0.54890        &-274                &7.27&4884&88\\
(2)$^3\Sigma^+$($^*$)&c    &          "          &     "          &"                   &8.25&\textit{+327}&73\\
  2$^{nd}$ well        &     &          "          &     "          &"                   &16.6&\textit{-1115}&22\\
   barrier           &     &         "            &     "          &"                   &11.0 &\textit{-1801}&- \\
(2)$^3\Pi$     &     &K$^+$+Sr($4s3d\,^3D$)&-0.52991        &343                 &8.64&3042&60\\
(1)$^3\Delta$  &     &          "          &     "          &"                   &8.26&3838&70\\
(3)$^3\Sigma^+$&     &          "          &     "          &"                   &11.0&1959&-\\
(1)$^1\Pi$     &B    &K$^+$+Sr($5s4d\,^1D$)&-0.52073        &460                 &8.34&4225&63\\
(3)$^1\Sigma^+$($^*$)&C    &          "          &     "          &"                   &8.53&\textit{-3757}&65\\
  2$^{nd}$ well       &     &          "          &     "          &"                   &15.2&\textit{-1714}&23\\
   barrier           &     &         "           &     "          &"                   &12.0 &\textit{-1603}&- \\
(1)$^1\Delta$  &     &          "          &     "          &"                   &8.18&5361&72\\
(2)$^1\Pi$     &D    &K$^+$+Sr($5s5p\,^1P$)&-0.51779        &-442                &8.32&2594&54\\
(4)$^1\Sigma^+$($^*$)&E    &          "          &     "          &"                   &13.1&\textit{1205}&48\\  \hline
\end{tabular}
\caption{Same as Table \ref{tab:constants_Li} for KSr$^+$. The $\omega_e$ value is not reported for the (3)$^3\Sigma^+$ curve, as it is not significant due to the peculiar shape of the well induced by the interaction with the (2)$^3\Sigma^+$ state.}
\label{tab:constants_K}
\end{table}
\begin{table}[t]
\center
\begin{tabular} {|c|c|c|c|c|c|c|c|} \hline
State          &Label& Diss. limit          &$E_{diss}$(a.u.)&$\delta$(cm$^{-1}$))&$R_e$&$D_e$& $\omega_e$\\ \hline
(1)$^1\Sigma^+$&X    &Rb$^+$+Sr($5s^2$)     &-0.61548        &-184                &8.49& 3638&55\\
(2)$^1\Sigma^+$&A    &Rb($5s$)+Sr$^+(5s)$   &-0.55886        &0                   &14.0& 848&21\\
(1)$^3\Sigma^+$&a    &          "           &     "          &"                   &9.41&6121&48\\
(1)$^3\Pi$     &b    &Rb$^+$+Sr($5s5p\,^3P$)&-0.54890        &-274                &7.90&3390&59\\
(2)$^3\Sigma^+$($^*$)&c    &          "           &     "          &"                   &8.83&\textit{+1492}&46\\
  2$^{nd}$ well        &     &          "           &     "          &"                   &18.0&\textit{-864}&15\\
   barrier           &     &         "            &     "          &"                   & 11.2&\textit{+2537}&- \\
(2)$^3\Pi$     &     &Rb$^+$+Sr($4s3d\,^3D$)&-0.52991        &343                 &9.32&2122&42\\
(1)$^3\Delta$  &     &          "           &     "          &"                   &8.90&2632&47\\
(3)$^3\Sigma^+$&     &          "           &     "          &"                   &11.3&1529&-\\
(1)$^1\Pi$     &B    &Rb$^+$+Sr($5s4d\,^1D$)&-0.52073        &460                 &9.0 &3056&43\\
(3)$^1\Sigma^+$($^*$)&C    &          "           &     "          &"                   &9.15&\textit{-2650}&44\\
  2$^{nd}$ well        &     &          "           &     "          &"                   &16.0&\textit{-1135}&17\\
   barrier           &     &         "            &     "          &"                   & 13.6&\textit{-941}&- \\
(1)$^1\Delta$  &     &          "           &     "          &"                   &8.77&3989&50\\
(2)$^1\Pi$     &D    &Rb$^+$+Sr($5s5p\,^1P$)&-0.51779        &-442                &9.18&1554&33\\
(4)$^1\Sigma^+$&E    &          "           &     "          &"                   &13.7& 925&-\\ \hline
\hline
\end{tabular}
\caption{Same as Table \ref{tab:constants_Li} for RbSr$^+$. The $\omega_e$ value is not reported for the (3)$^3\Sigma^+$ and (4)$^1\Sigma^+$ curves, as it is not significant, due to their peculiar shape.}
\label{tab:constants_Rb}
\end{table}
\begin{table}[t]
\center
\begin{tabular} {|c|c|c|c|c|c|c|c|} \hline
State          &Label& Diss. limit          &$E_{diss}$(a.u.)&$\delta$(cm$^{-1}$))&$R_e$&$D_e$& $\omega_e$\\ \hline
(1)$^1\Sigma^+$&X    &Cs$^+$+Sr($5s^2$)     &-0.61548        &-184                &9.05&2714&44\\
(2)$^1\Sigma^+$&A    &Cs($6s$)+Sr$^+(5s)$   &-0.54845        &0                   &13.9&1133&20\\
(1)$^3\Sigma^+$&a    &         "            &     "          &"                   &9.90&7003&41\\
(1)$^3\Pi$     &b    &Cs$^+$+Sr($5s5p\,^3P$)&-0.54890        &-274                &8.30&2738&49\\
(2)$^3\Sigma^+$($^*$)&c    &         "            &     "          &"                   &9.38&\textit{+2293}&37\\
  2$^{nd}$ well        &     &         "            &     "          &"                   &22.5&\textit{-62}&5\\
   barrier           &     &         "            &     "          &"                   & 12.0&\textit{-1133}&  \\
(2)$^3\Pi$     &     &Cs$^+$+Sr($4s3d\,^3D$)&-0.52991        &343                 &9.90&1450&33\\
(1)$^3\Delta$  &     &         "            &     "          &"                   &9.46&1779&37\\
(3)$^3\Sigma^+$&     &         "            &     "          &"                   &12.4& 794&-\\
(1)$^1\Pi$     &B    &Cs$^+$+Sr($5s4d\,^1D$)&-0.52073        &460                 &9.41&2358&33\\
(3)$^1\Sigma^+$($^*$)&C    &         "            &     "          &"                   &9.67&\textit{-1874}&35\\
  2$^{nd}$ well        &     &         "            &     "          &"                   &17.0&\textit{-784} &13\\
   barrier           &     &         "            &     "          &"                   & 14.2&\textit{-607}&  \\
(1)$^1\Delta$  &     &         "            &     "          &"                   &9.29&2987&41\\
(2)$^1\Pi$     &D    &Cs$^+$+Sr($5s5p\,^1P$)&-0.51779        &-442                &9.67&1277&30\\
(4)$^1\Sigma^+$&E    &         "            &     "          &"                   &14.6& 597&58\\\hline
\end{tabular}
\caption{Same as Table \ref{tab:constants_Li} for CsSr$^+$. The $\omega_e$ value is not reported for the (3)$^3\Sigma^+$ curve, as it is not significant, due to the peculiar shape of the well induced by the interaction with the (2)$^3\Sigma^+$ state.}
\label{tab:constants_Cs}
\end{table}
\section{Permanent and transition dipole moments, static dipole polarizabilities}
\label{sec:other}
\begin{figure}[htb]
  \begin{center}
    \includegraphics[width=0.6\textwidth]{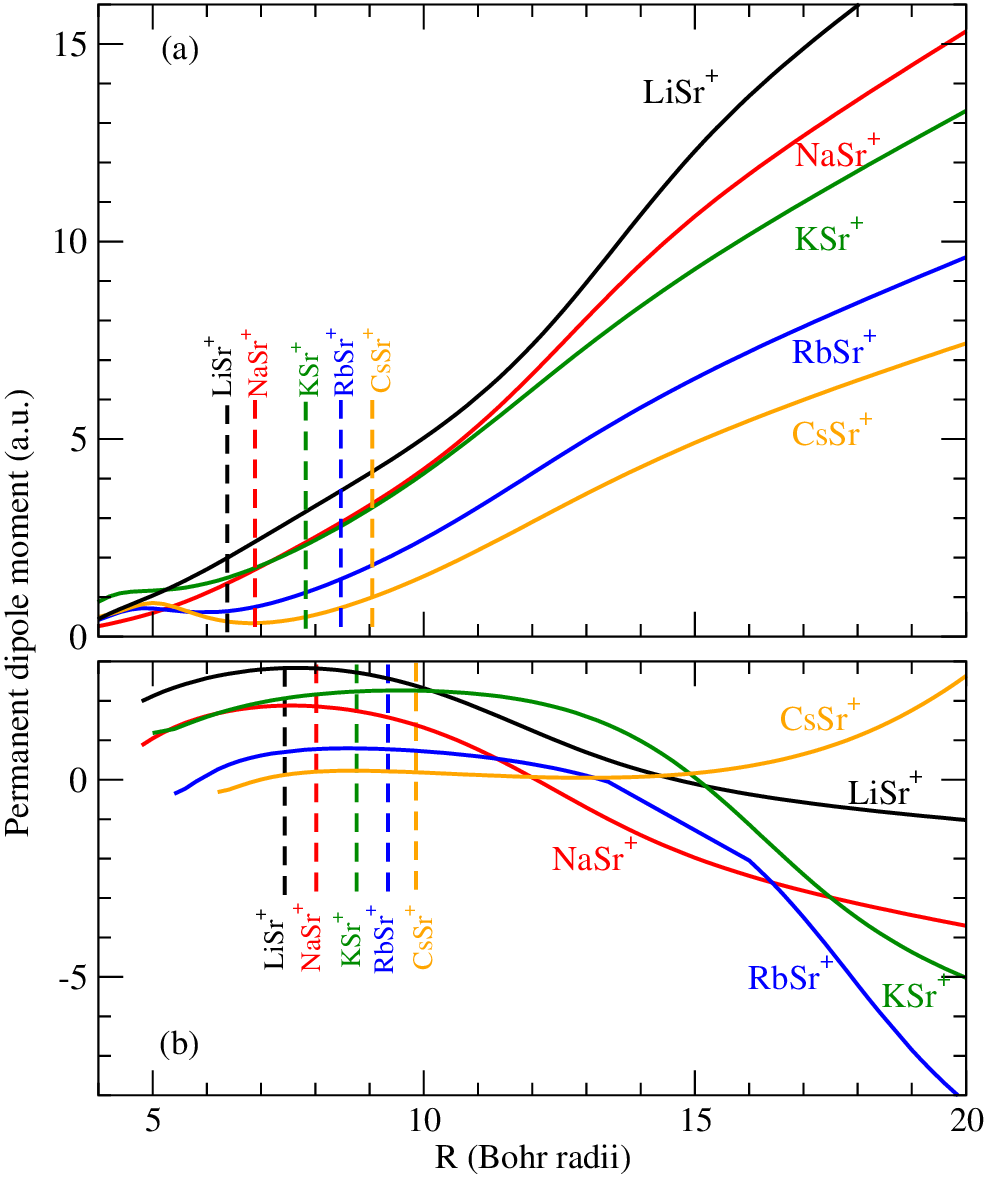}
  \end{center}
  \caption{Permanent dipole moments (a.u.) of (a) the X$^1\Sigma^+$ ground state and (b) the a$^3\Sigma^+$ metastable state of the $A$-Sr$^+$ molecules computed in the present work. The origin is taken at the center-of-mass of the molecule, considering the $^{87}$Sr, $^{6}$Li, and $^{85}$Rb isotopes. The position of equilibrium distances are shown with vertical dashed lines.}
  \label{fig:pdm}
\end{figure}

\begin{figure}[htb]
  \begin{center}
    \includegraphics[width=0.6\textwidth]{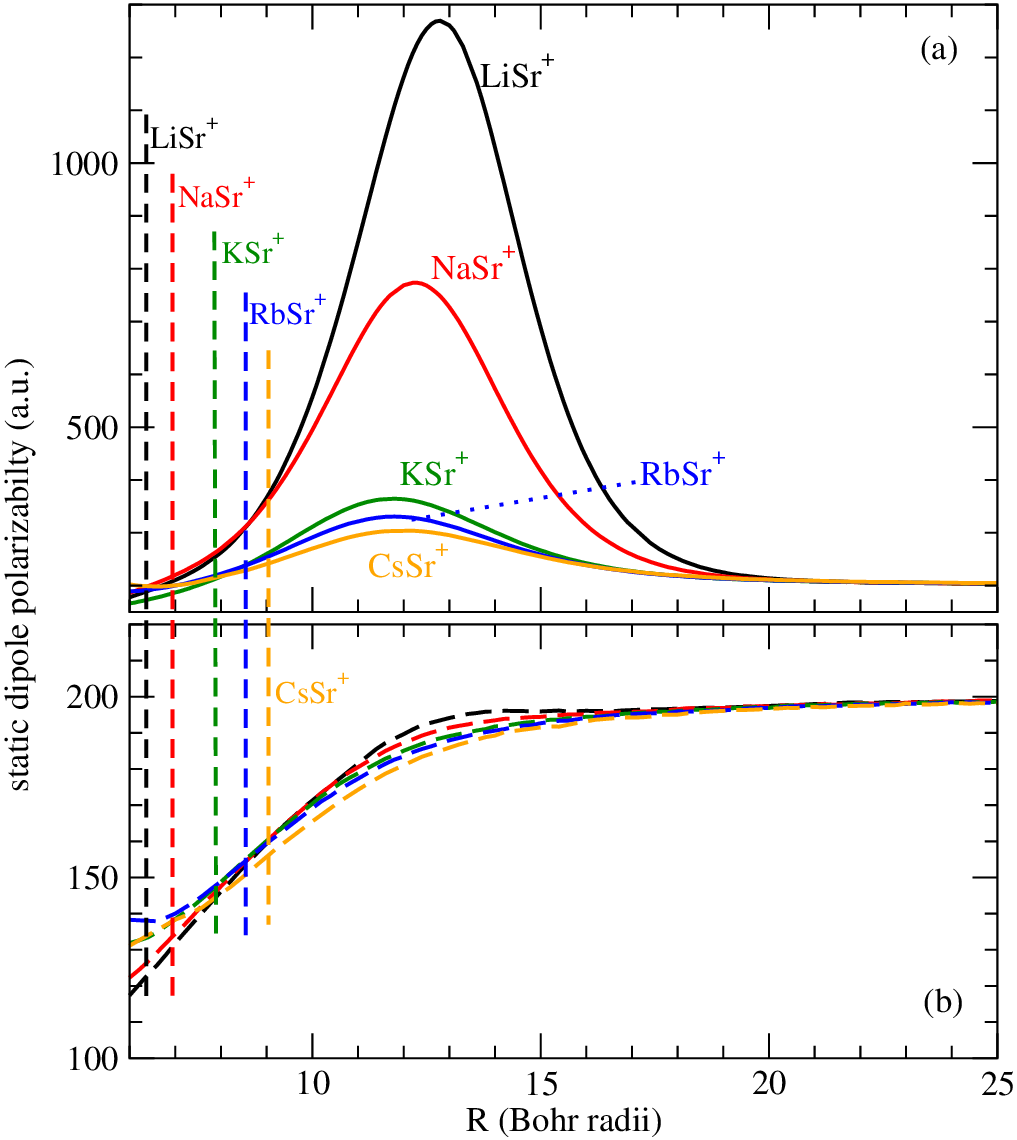}
  \end{center}
  \caption{(a) Parallel and (b) perpendicular components of the static dipole polarizability (in a.u., 1~a.u.=0.148 184 709 3~\AA $^3$) of the X$^1\Sigma^+$ ground state of the $A$-Sr$^+$ molecules computed in the present work. The position of equilibrium distances are shown with vertical dashed lines.}
  \label{fig:pola}
\end{figure}

\begin{figure}[htb]
  \begin{center}
    \includegraphics[width=0.6\textwidth]{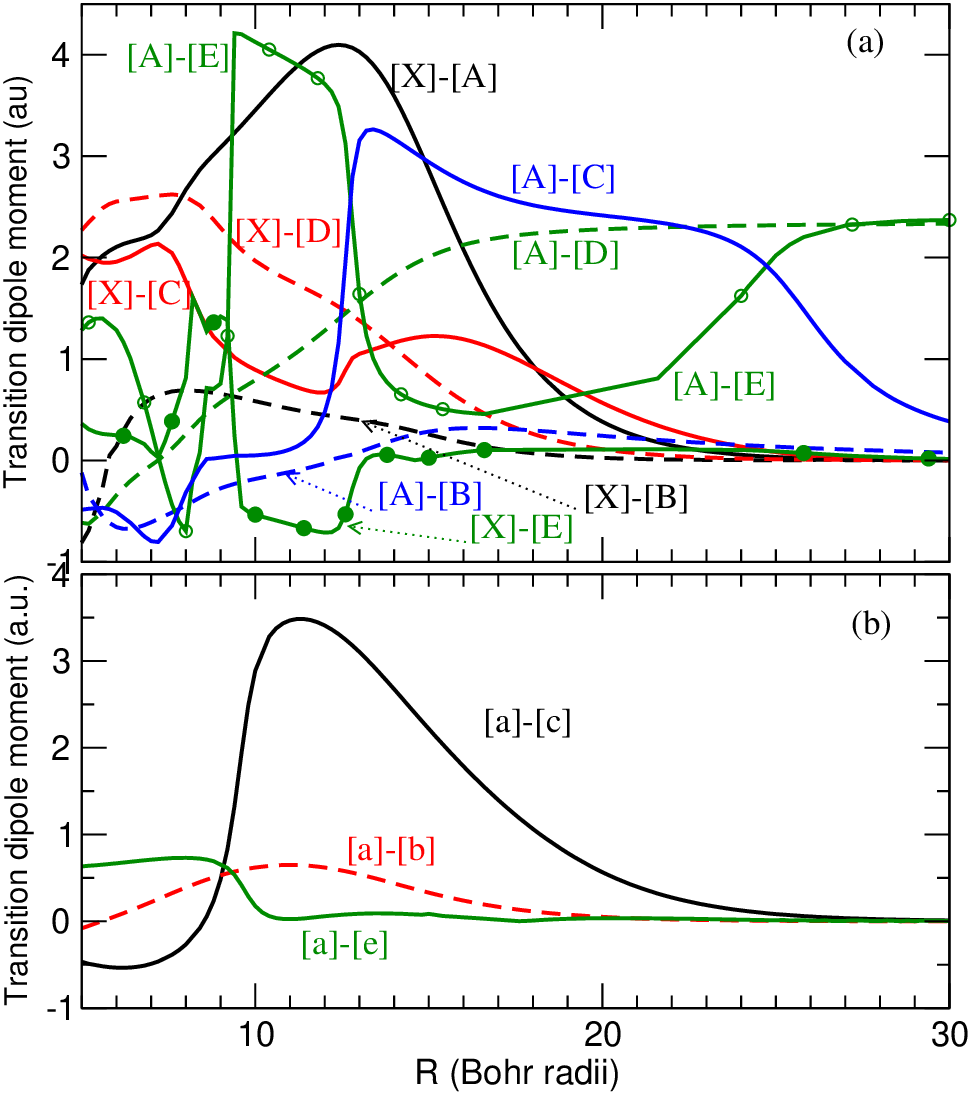}
  \end{center}
  \caption{Selected transition dipole moments between (a) singlet states, and (b) triplet states, for LiSr$^+$. Molecular states are labeled with letters as in Table \ref{tab:constants_Li}. }
  \label{fig:tdm_lisr+}
\end{figure}

\begin{figure}[htb]
  \begin{center}
    \includegraphics[width=0.6\textwidth]{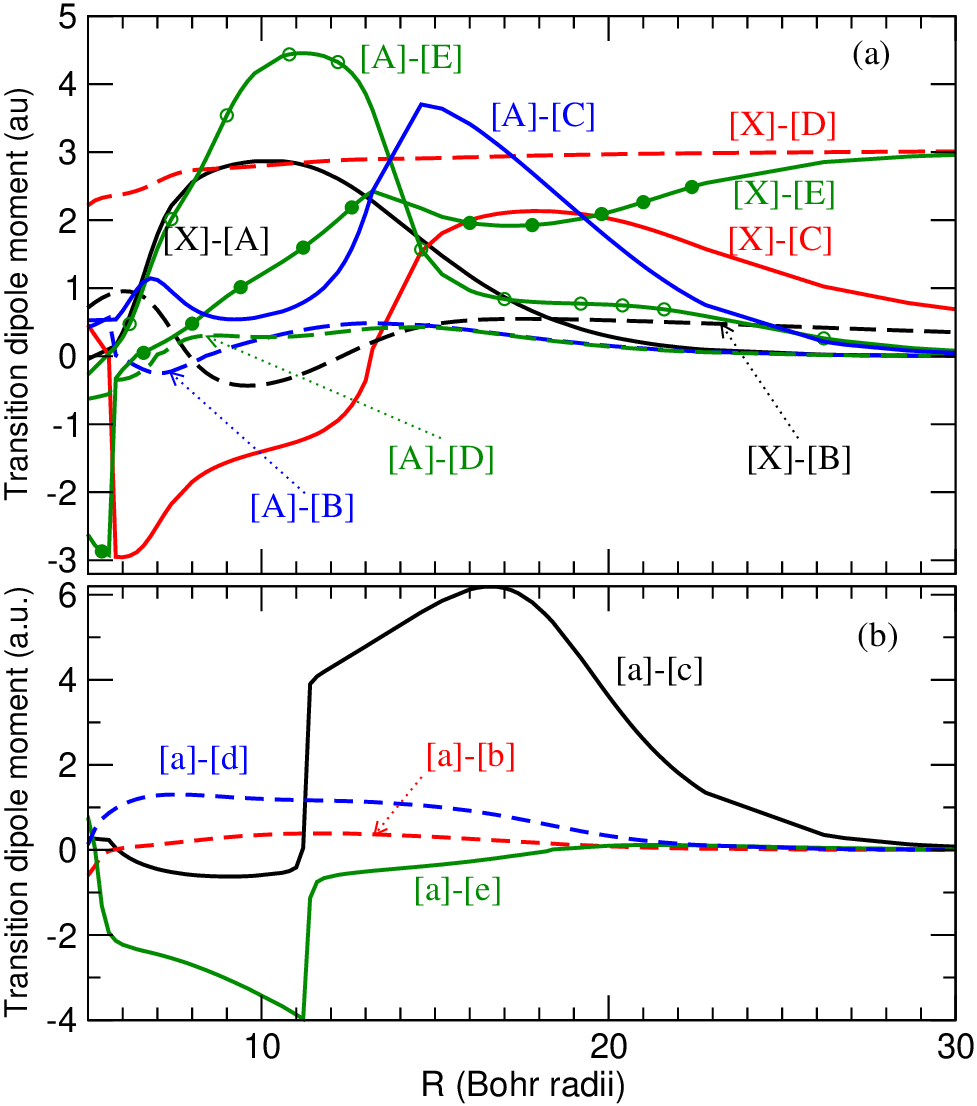}
  \end{center}
  \caption{Selected transition dipole moments between (a) singlet states, and (b) triplet states, for RbSr$^+$. Molecular states are labeled with letters as in Table \ref{tab:constants_Rb}. }
  \label{fig:tdm_rbsr+}
\end{figure}

Permanent dipole moments (PDM) are relevant to the present study, \textit{i.e.} if molecular ions in the X$^1\Sigma^+$ ground state or in the a$^3\Sigma^+$ metastable state are created in experiment manipulating cold Sr$^+$ and alkali-metal atoms: PDMs will determine the ability to perform vibrational or rotational cooling following for instance methods recently demonstrated for MgH$^+$ \cite{staanum2010} and HD$^+$ \cite{schneider2010}. The PDM is here calculated in the body-fixed frame with respect to the origin placed at the center-of-mass of the molecule, and is therefore characterized by a linear divergence at large internuclear distances expressing the increasing distance between the negative and positive center-of-charge. Our results are reported in Fig.\ref{fig:pdm} for these two electronic states and for all $A$-Sr$^+$ species, confirming this pattern. The molecular axis is oriented along the Sr$^+$-$A$ direction, so that the positive sign of the ground state PDM indicates an excess of electron charge on the alkali-metal atom, which is less pronounced as its mass increases. This is compatible with the decrease of the electron affinity of alkali-metal species with their increasing mass. The PDM is almost linear in the region of the minimum of the ground state, so that radiative transitions among vibrational levels levels should be possible. Its magnitude is also significant, so that rotational transitions should be observed. In contrast the sign of the triplet PDM, while mostly positive, cannot be easily interpreted over the whole distance range. Note that its dissociation limit is different from the ground state one, so that its divergence at large distances is different. Its magnitude is almost constant (and even close to zero for CsSr$^+$), which therefore is probably not a good candidate for cooling of internal degrees of freedom.
As a complement to the description of the properties of the $A$-Sr$^+$ molecules, we computed their static dipole polarizabilities (SDP), which will be relevant for instance when considering the long-range interaction with surrounding cold neutral alkali-metal atoms in a merged cold ion/atom trap experiment. As pointed out in Ref.\cite{sternheimer1969}, the SDP includes a core contribution (5.67~a.u. here, see Table II of paper I) in addition to the valence contribution, the latter being evaluated using the finite field method \cite{cohen1965}. Both parallel $\alpha_{\parallel}$ (along the internuclear axis), and perpendicular $\alpha_{\perp}$ components (see Fig.\ref{fig:pola}) exhibit  an $R$-variation similar to the one obtained for alkali dimers \cite{deiglmayr2008}, alkali hydrides  \cite{aymar2009}, for MgH$^+$ \cite{aymar2009a}, or for the hydrogen molecule \cite{kolos1967}. In all these systems, $\alpha_{\parallel}$ has a maximum at a distance larger than the equilibrium distance $R_e$ of the system, while $\alpha_{\perp}$ always has a smaller  magnitude than $\alpha_{\parallel}$, and monotonically increases towards the asymptotic limit. Around the equilibrium distance, $\alpha_{\parallel}$ ranges between 200 and 250~a.u., and $\alpha_{\perp}$ between 120 and 150~a.u.. As for the well depth the maximum of $\alpha_{\parallel}$ increases as the mass of the $A$-Sr$^+$ molecule decreases. At large distances, both components converge toward a unique limit corresponding to the sum of SDP of the Sr ground state and the (small) SDP of the alkali-metal ions.
Finally, the electronic transition dipole moments (TDM) are relevant when considering for instance the possibility for radiative charge exchange (RCE) in these systems \cite{makarov2003,idziaszek2009}, or for photoassociation (PA) and spontaneous emission (SE) (see next section). We have computed all TDMs of all $A$-Sr$^+$ species, whose numerical data are provided through the EPAPS system. In order to illustrate them, we chose the LiSr$^+$ and the RbSr$^+$ molecules as being representative of the $A$-Sr$^+$ species (see Section \ref{sec:pot}). We represent in Figs. \ref{fig:tdm_lisr+} and \ref{fig:tdm_rbsr+} the TDMs involving $X^1\Sigma^+$, (2)$^1\Sigma^+$ and $a^3\Sigma^+$ as the lowest states. For the sake of clarity, we will use both numbered (in parenthesis) and lettered (when available, in brackets) notations for the molecular states. Various patterns can be analyzed in these figures.

The most obvious one is when the TDM function connects two molecular states which dissociate towards atomic states connected by a resonant transition, as for the $(1)^1\Sigma^+[X] - (2)^1\Pi[D]$ and $(1)^1\Sigma^+[X] - (4)^1\Sigma^+[E]$ in RbSr$^+$: the TDM function converges towards the Sr$^+$ ($^1S-^1P$) atomic TDM. The $R$-variation of the former suggests that the molecular states $[X]$ and $[D]$ keep a similar electronic character, while the abrupt variation of the latter reflects the strong configuration interaction of the $[E]$ state, induced by the avoided crossing with the $(3)^1\Sigma^+[C]$ state. It is actually clear on Fig.\ref{fig:tdm_rbsr+}(a) that the $[C]$ and $[E]$ state exchange their electronic character at the position of their avoided crossing around 14~a.u.. The other displayed TDM functions for  $(1)^1\Sigma^+[X] - (2)^1\Sigma^+[A]$, $(1)^1\Sigma^+[X] - (1)^1\Pi[B]$, and $(1)^1\Sigma^+[X] - (3)^1\Sigma^+[C]$ vanishes at large distances, with strong variations at short internuclear distances again related to strong configuration interaction. The same analysis holds for transitions involving the $[A]$ state. In particular, the states connected to $[A]$, \textit{i.e.} $[C]$, $[D]$, and $[E]$ have vanishing TDMs at large distances, as they correspond asymptotically to different charge states of the separated atoms. A similar reasoning can be made for LiSr$^+$, where the results reflect the different ordering of the asymptotic limits compared to the other species. For both species represented in \ref{fig:tdm_lisr+}b and \ref{fig:tdm_rbsr+}b, the TDM functions for the $(1)^3\Sigma^+[a] - (2)^3\Sigma^+[c]$ and $(1)^3\Sigma^+[a] - (3)^3\Sigma^+[e]$ exchange their character at the distance of the avoided crossing between the $[c]$ and $[e]$ states around 9~a.u. and 11~a.u. for LiSr$^+$ and RbSr$^+$, respectively.

\section{Discussion: radiative charge exchange, photoassociation, and stability of $A$-Sr$^+$ molecules }
\label{sec:discussion}

The experiments aiming at merging a cold Sr$^+$ trap and a cold alkali-metal atom trap involve molecular states connected to the second dissociation limit, \textit{i.e.} the $(2)^1\Sigma^+[A]$ and the $(1)^3\Sigma^+[a]$ states. Despite the strong variation of the $[X]-[A]$ TDM between 10 and 20a.u., it is most probable that the RCE process during a cold collision between ground state alkali-metal atom $A$ and Sr$^+$ will be weak as the TDM rapidly vanishes with $R$. This is actually the trend already reported in Ref.\cite{makarov2003} for Na-Ca$^+$ collisions. However, as it is possible to trap clouds of laser-cooled Sr$^+$ ions for hours, it may be possible to observe the slow decreasing of Sr$^+$ ions, while the $A^+$ alkali-metal ions will manifest as dark spots in the ion crystal. The trapped ions are continuously illuminated by the cooling lasers, so that excited alkali-metal atom $A^*$ in their lowest $p$ state are present. In this perspective, the Li-Sr$^+$ exhibits a peculiar situation: as noticed in Section \ref{sec:pot} the Li($2p$)+Sr$^+$(5s) limit lies only 178~cm$^{-1}$ above the Li($2s$)+Sr$^+(4d)$ one. The  transition between the $[D]$ and $[E]$ $^1\Sigma^+$ enhanced by their mutual avoided crossing may lead to a strong excitation exchange, leaving Sr$^+$ ions in their $4d$ state, insensitive to the cooling laser frequency. More generally, the asymptotes involving the resonantly excited Sr$^+(5p)$ lie outside of the energy range of the presented figures, where a large number of potential curves probably offer many channels for charge exchange. It will be interesting to check the variation with the intensity of the cooling lasers of the ion number in such experiments, which would require an extension of the present treatment.

Prospects for efficient PA of $A$-Sr$^+$ pairs, and stabilization into cold and trapped $A$Sr$^+$ molecular ions look promising, according to Figs \ref{fig:Li} and \ref{fig:Rb}. The Li-Sr$^+$ and Rb-Sr$^+$ molecules are taken as representative examples, due to the above mentioned peculiarity of Li-Sr$^+$, and the use of Rb atoms in several merged ion trap/cold atom trap experiments. For instance, the excitation of the $(2)^1\Pi$ state from the entrance channel Li$(2s)$+Sr$^+(5s)$ with a wavelength slightly larger than the one of the $2s-2P$ transition in Li could take place either from large distances, or from the inner turning point of the $(2)^1\Sigma^+$ potential curve. It could lead to an efficient decay down to the low-lying levels of the ground state. A similar mechanism is predicted for the formation of metastable $(1)^3\Sigma^+$ ions in deeply-bound levels via the excitation of the $(3)^3\Sigma^+$. PA of Rb and Sr$^+$ pairs are expected using a wavelength detuned to the red of the $5s^2-5s5p^1P$ Sr$^+$ resonance. An efficient decay is expected to take place from the inner turning point of the $(4)^1\Sigma^+$. Another path could be the excitation at the turning point of the $(2)^1\Sigma^+$ toward the $(2)^1\Pi$, which would be stabilized into the lowest ground state levels. No triplet states will ba accessible in such a scheme. But the spin-orbit coupling have been neglected so far. It is quite large in the Sr$^+$ ion (280.34~cm$^{-1}$ for the $4d$ level, and 801.46~cm$^{-1}$ for the $5p$ level \cite{moore1958}), and comparable in some cases with the one of alkali-metal atoms (0.29, 17.2, 59.7, 237.9, and 551.3~cm$^{-1}$, for Li, Na, K, Rb, and Cs, respectively). This will induce additional couplings between molecular states of various symmetries. For instance, the spin-orbit interaction will strongly mix the $(2)^1\Pi$ and the $(3)^3\Sigma^+$, probably leading to strong predissociation channels with charge exchange into the Rb$^+$+Sr($^3P,^3D$) outgoing channels. A detailed description would require further quantum chemistry calculations which are beyond the scope of the present paper.

Once $A$Sr$^+$ ions are formed, the important issue of their stability against charge exchange with the surrounding cold alkali-metal atoms is raised, just like in the proposal of Ref.\cite{hudson2009} looking at possible sympathetic cooling of molecular ions by laser-cooled atoms. In our case, the ionization potential (IP) of alkali-metal atoms should be larger than the IP of the related neutral molecule. Combining the present results with those of paper I for the ground state of neutral $A$Sr molecules, we estimate their IP by the energy difference between the minimum $D_e$ of their ground state potential curves according to: IP($A$Sr)=IP(A)+$D_e$($A$Sr)-$D_e$($A$Sr$^+$). Using the experimental IP for alkali-metal atoms \cite{moore1958}, namely 43487, 41449, 35009, 33691, and 31046~cm$^{-1}$ for Li, Na, K, Rb, and Cs respectively, we find IP(LiSr)=34848~cm$^{-1}$, IP(NaSr)=34716~cm$^{-1}$, IP(KSr)=31198~cm$^{-1}$, IP(RbSr)=31126~cm$^{-1}$, IP(CsSr)=29416~cm$^{-1}$. Therefore we predict that  the $A$Sr$^+$ ions will be all stable against charge exchange with alkali-metal atom $A$.

\section*{Acknowledgments}
R.G. gratefully acknowledges support from Institut Francilien de Recherches sur les Atomes Froids. This work has
been performed in the framework of the network “Quantum Gases of Dipolar Molecules” of the EUROQUAM program of
the European Science Foundation (ESF).

\newpage
\section*{References}


\begin{thebibliography}{10}%
\makeatletter
\providecommand \@ifxundefined [1]{%
 \ifx #1\undefined \expandafter \@firstoftwo
 \else \expandafter \@secondoftwo
\fi
}%
\providecommand \@ifnum [1]{%
 \ifnum #1\expandafter \@firstoftwo
 \else \expandafter \@secondoftwo
\fi
}%
\providecommand \enquote [1]{``#1''}%
\providecommand \bibnamefont  [1]{#1}%
\providecommand \bibfnamefont [1]{#1}%
\providecommand \citenamefont [1]{#1}%
\providecommand\href[0]{\@sanitize\@href}%
\providecommand\@href[1]{\endgroup\@@startlink{#1}\endgroup\@@href}%
\providecommand\@@href[1]{#1\@@endlink}%
\providecommand \@sanitize [0]{\begingroup\catcode`\&12\catcode`\#12\relax}%
\@ifxundefined \pdfoutput {\@firstoftwo}{%
 \@ifnum{\z@=\pdfoutput}{\@firstoftwo}{\@secondoftwo}%
}{%
 \providecommand\@@startlink[1]{\leavevmode}%
 \providecommand\@@endlink[0]{}%
}{%
 \providecommand\@@startlink[1]{%
  \leavevmode
  \pdfstartlink
   attr{/Border[0 0 1 ]/H/I/C[0 1 1]}%
   user{/Subtype/Link/A<</Type/Action/S/URI/URI(#1)>>}%
  \relax
 }%
 \providecommand\@@endlink[0]{\pdfendlink}%
}%
\providecommand \url  [0]{\begingroup\@sanitize \@url }%
\providecommand \@url [1]{\endgroup\@href {#1}{\urlprefix}}%
\providecommand \urlprefix [0]{URL }%
\providecommand \Eprint[0]{\href }%
\@ifxundefined \urlstyle {%
  \providecommand \doi [1]{doi:\discretionary{}{}{}#1}%
}{%
  \providecommand \doi [0]{doi:\discretionary{}{}{}\begingroup
  \urlstyle{rm}\Url }%
}%
\providecommand \doibase [0]{http://dx.doi.org/}%
\providecommand \Doi[1]{\href{\doibase#1}}%
\providecommand \bibAnnote [3]{%
  \BibitemShut{#1}%
  \begin{quotation}\noindent
    \textsc{Key:}\ #2\\\textsc{Annotation:}\ #3%
  \end{quotation}%
}%
\providecommand \bibAnnoteFile [2]{%
  \IfFileExists{#2}{\bibAnnote {#1} {#2} {\input{#2}}}{}%
}%
\providecommand \typeout [0]{\immediate \write \m@ne }%
\providecommand \selectlanguage [0]{\@gobble}%
\providecommand \bibinfo [0]{\@secondoftwo}%
\providecommand \bibfield [0]{\@secondoftwo}%
\providecommand \translation [1]{[#1]}%
\providecommand \BibitemOpen[0]{}%
\providecommand \bibitemStop [0]{}%
\providecommand \bibitemNoStop [0]{.\EOS\space}%
\providecommand \EOS [0]{\spacefactor3000\relax}%
\providecommand \BibitemShut [1]{\csname bibitem#1\endcsname}%
\bibitem{carr2009}%
  \BibitemOpen
  \bibfield{author}{%
  \bibinfo {author} {\bibfnamefont{L.~D.}\ \bibnamefont{Carr}}\ and\ \bibinfo
  {author} {\bibfnamefont{J.}~\bibnamefont{Ye}},\ }%
  \bibfield{journal}{%
  \bibinfo {journal} {New J. Phys.}\ }%
  \textbf{\bibinfo {volume} {11}},\ \bibinfo {pages} {055009} (\bibinfo {year}
  {2009})%
  \bibAnnoteFile{NoStop}{carr2009}%
\bibitem{dulieu2009}%
  \BibitemOpen
  \bibfield{author}{%
  \bibinfo {author} {\bibfnamefont{O.}~\bibnamefont{Dulieu}}\ and\ \bibinfo
  {author} {\bibfnamefont{C.}~\bibnamefont{Gabbanini}},\ }%
  \bibfield{journal}{%
  \bibinfo {journal} {Rep. Prog. Phys.}\ }%
  \textbf{\bibinfo {volume} {72}},\ \bibinfo {pages} {086401} (\bibinfo {year}
  {2009})%
  \bibAnnoteFile{NoStop}{dulieu2009}%
\bibitem{fioretti1998}%
  \BibitemOpen
  \bibfield{author}{%
  \bibinfo {author} {\bibfnamefont{A.}~\bibnamefont{Fioretti}}, \bibinfo
  {author} {\bibfnamefont{D.}~\bibnamefont{Comparat}}, \bibinfo {author}
  {\bibfnamefont{A.}~\bibnamefont{Crubellier}}, \bibinfo {author}
  {\bibfnamefont{O.}~\bibnamefont{Dulieu}}, \bibinfo {author}
  {\bibfnamefont{F.}~\bibnamefont{Masnou-Seeuws}},\ and\ \bibinfo {author}
  {\bibfnamefont{P.}~\bibnamefont{Pillet}},\ }%
  \bibfield{journal}{%
  \bibinfo {journal} {Phys. Rev. Lett.}\ }%
  \textbf{\bibinfo {volume} {80}},\ \bibinfo {pages} {4402} (\bibinfo {year}
  {1998})%
  \bibAnnoteFile{NoStop}{fioretti1998}%
\bibitem{takekoshi1998}%
  \BibitemOpen
  \bibfield{author}{%
  \bibinfo {author} {\bibfnamefont{T.}~\bibnamefont{Takekoshi}}, \bibinfo
  {author} {\bibfnamefont{B.~M.}\ \bibnamefont{Patterson}},\ and\ \bibinfo
  {author} {\bibfnamefont{R.~J.}\ \bibnamefont{Knize}},\ }%
  \bibfield{journal}{%
  \bibinfo {journal} {Phys. Rev. Lett.}\ }%
  \textbf{\bibinfo {volume} {81}},\ \bibinfo {pages} {5105} (\bibinfo {year}
  {1998})%
  \bibAnnoteFile{NoStop}{takekoshi1998}%
\bibitem{takekoshi1999}%
  \BibitemOpen
  \bibfield{author}{%
  \bibinfo {author} {\bibfnamefont{T.}~\bibnamefont{Takekoshi}}, \bibinfo
  {author} {\bibfnamefont{B.~M.}\ \bibnamefont{Patterson}},\ and\ \bibinfo
  {author} {\bibfnamefont{R.~J.}\ \bibnamefont{Knize}},\ }%
  \bibfield{journal}{%
  \bibinfo {journal} {Phys. Rev. A}\ }%
  \textbf{\bibinfo {volume} {59}},\ \bibinfo {pages} {R5} (\bibinfo {year}
  {1999})%
  \bibAnnoteFile{NoStop}{takekoshi1999}%
\bibitem{nikolov2000}%
  \BibitemOpen
  \bibfield{author}{%
  \bibinfo {author} {\bibfnamefont{A.~N.}\ \bibnamefont{Nikolov}}, \bibinfo
  {author} {\bibfnamefont{J.~R.}\ \bibnamefont{Enscher}}, \bibinfo {author}
  {\bibfnamefont{E.~E.}\ \bibnamefont{Eyler}}, \bibinfo {author}
  {\bibfnamefont{H.}~\bibnamefont{Wang}}, \bibinfo {author}
  {\bibfnamefont{W.~C.}\ \bibnamefont{Stwalley}},\ and\ \bibinfo {author}
  {\bibfnamefont{P.~L.}\ \bibnamefont{Gould}},\ }%
  \bibfield{journal}{%
  \bibinfo {journal} {Phys. Rev. Lett.}\ }%
  \textbf{\bibinfo {volume} {84}},\ \bibinfo {pages} {246} (\bibinfo {year}
  {2000})%
  \bibAnnoteFile{NoStop}{nikolov2000}%
\bibitem{sage2005}%
  \BibitemOpen
  \bibfield{author}{%
  \bibinfo {author} {\bibfnamefont{J.~M.}\ \bibnamefont{Sage}}, \bibinfo
  {author} {\bibfnamefont{S.}~\bibnamefont{Sainis}}, \bibinfo {author}
  {\bibfnamefont{T.}~\bibnamefont{Bergeman}},\ and\ \bibinfo {author}
  {\bibfnamefont{D.}~\bibnamefont{DeMille}},\ }%
  \bibfield{journal}{%
  \bibinfo {journal} {Phys. Rev. Lett.}\ }%
  \textbf{\bibinfo {volume} {94}},\ \bibinfo {pages} {203001} (\bibinfo {year}
  {2005})%
  \bibAnnoteFile{NoStop}{sage2005}%
\bibitem{deiglmayr2008a}%
  \BibitemOpen
  \bibfield{author}{%
  \bibinfo {author} {\bibfnamefont{J.}~\bibnamefont{Deiglmayr}}, \bibinfo
  {author} {\bibfnamefont{A.}~\bibnamefont{Grochola}}, \bibinfo {author}
  {\bibfnamefont{M.}~\bibnamefont{Repp}}, \bibinfo {author}
  {\bibfnamefont{K.}~\bibnamefont{M\"ortlbauer}}, \bibinfo {author}
  {\bibfnamefont{C.}~\bibnamefont{Gl\"uck}}, \bibinfo {author}
  {\bibfnamefont{J.}~\bibnamefont{Lange}}, \bibinfo {author}
  {\bibfnamefont{O.}~\bibnamefont{Dulieu}}, \bibinfo {author}
  {\bibfnamefont{R.}~\bibnamefont{Wester}},\ and\ \bibinfo {author}
  {\bibfnamefont{M.}~\bibnamefont{Weidem\"uller}},\ }%
  \bibfield{journal}{%
  \bibinfo {journal} {Phys. Rev. Lett.}\ }%
  \textbf{\bibinfo {volume} {101}},\ \bibinfo {pages} {133004} (\bibinfo {year}
  {2008})%
  \bibAnnoteFile{NoStop}{deiglmayr2008a}%
\bibitem{viteau2008}%
  \BibitemOpen
  \bibfield{author}{%
  \bibinfo {author} {\bibfnamefont{M.}~\bibnamefont{Viteau}}, \bibinfo {author}
  {\bibfnamefont{A.}~\bibnamefont{Chotia}}, \bibinfo {author}
  {\bibfnamefont{M.}~\bibnamefont{Allegrini}}, \bibinfo {author}
  {\bibfnamefont{N.}~\bibnamefont{Bouloufa}}, \bibinfo {author}
  {\bibfnamefont{O.}~\bibnamefont{Dulieu}}, \bibinfo {author}
  {\bibfnamefont{D.}~\bibnamefont{Comparat}},\ and\ \bibinfo {author}
  {\bibfnamefont{P.}~\bibnamefont{Pillet}},\ }%
  \bibfield{journal}{%
  \bibinfo {journal} {Science}\ }%
  \textbf{\bibinfo {volume} {321}},\ \bibinfo {pages} {232} (\bibinfo {year}
  {2008})%
  \bibAnnoteFile{NoStop}{viteau2008}%
\bibitem{ni2008}%
  \BibitemOpen
  \bibfield{author}{%
  \bibinfo {author} {\bibfnamefont{K.-K.}\ \bibnamefont{Ni}}, \bibinfo {author}
  {\bibfnamefont{S.}~\bibnamefont{Ospelkaus}}, \bibinfo {author}
  {\bibfnamefont{M.~H.~G.}\ \bibnamefont{de~Miranda}}, \bibinfo {author}
  {\bibfnamefont{A.}~\bibnamefont{Peer}}, \bibinfo {author}
  {\bibfnamefont{B.}~\bibnamefont{Neyenhuis}}, \bibinfo {author}
  {\bibfnamefont{J.~J.}\ \bibnamefont{Zirbel}}, \bibinfo {author}
  {\bibfnamefont{S.}~\bibnamefont{Kotochigova}}, \bibinfo {author}
  {\bibfnamefont{P.~S.}\ \bibnamefont{Julienne}}, \bibinfo {author}
  {\bibfnamefont{D.~S.}\ \bibnamefont{Jin}},\ and\ \bibinfo {author}
  {\bibfnamefont{J.}~\bibnamefont{Ye}},\ }%
  \bibfield{journal}{%
  \bibinfo {journal} {Science}\ }%
  \textbf{\bibinfo {volume} {322}},\ \bibinfo {pages} {231} (\bibinfo {year}
  {2008})%
  \bibAnnoteFile{NoStop}{ni2008}%
\bibitem{danzl2008}%
  \BibitemOpen
  \bibfield{author}{%
  \bibinfo {author} {\bibfnamefont{J.~G.}\ \bibnamefont{Danzl}}, \bibinfo
  {author} {\bibfnamefont{E.}~\bibnamefont{Haller}}, \bibinfo {author}
  {\bibfnamefont{M.}~\bibnamefont{Gustavsson}}, \bibinfo {author}
  {\bibfnamefont{M.~J.}\ \bibnamefont{Mark}}, \bibinfo {author}
  {\bibfnamefont{R.}~\bibnamefont{Hart}}, \bibinfo {author}
  {\bibfnamefont{N.}~\bibnamefont{Bouloufa}}, \bibinfo {author}
  {\bibfnamefont{O.}~\bibnamefont{Dulieu}}, \bibinfo {author}
  {\bibfnamefont{H.}~\bibnamefont{Ritsch}},\ and\ \bibinfo {author}
  {\bibfnamefont{H.-C.}\ \bibnamefont{N\"{a}gerl}},\ }%
  \bibfield{journal}{%
  \bibinfo {journal} {Science}\ }%
  \textbf{\bibinfo {volume} {321}},\ \bibinfo {pages} {1062} (\bibinfo {year}
  {2008})%
  \bibAnnoteFile{NoStop}{danzl2008}%
\bibitem{lang2008a}%
  \BibitemOpen
  \bibfield{author}{%
  \bibinfo {author} {\bibfnamefont{F.}~\bibnamefont{Lang}}, \bibinfo {author}
  {\bibfnamefont{K.}~\bibnamefont{Winkler}}, \bibinfo {author}
  {\bibfnamefont{C.}~\bibnamefont{Strauss}}, \bibinfo {author}
  {\bibfnamefont{R.}~\bibnamefont{Grimm}},\ and\ \bibinfo {author}
  {\bibfnamefont{J.~H.}\ \bibnamefont{Denschlag}},\ }%
  \bibfield{journal}{%
  \bibinfo {journal} {Phys. Rev. Lett.}\ }%
  \textbf{\bibinfo {volume} {101}},\ \bibinfo {pages} {133005} (\bibinfo {year}
  {2008})%
  \bibAnnoteFile{NoStop}{lang2008a}%
\bibitem{danzl2010}%
  \BibitemOpen
  \bibfield{author}{%
  \bibinfo {author} {\bibfnamefont{J.~G.}\ \bibnamefont{Danzl}}, \bibinfo
  {author} {\bibfnamefont{M.~J.}\ \bibnamefont{Mark}}, \bibinfo {author}
  {\bibfnamefont{E.}~\bibnamefont{Haller}}, \bibinfo {author}
  {\bibfnamefont{M.}~\bibnamefont{Gustavsson}}, \bibinfo {author}
  {\bibfnamefont{R.}~\bibnamefont{Hart}}, \bibinfo {author}
  {\bibfnamefont{J.}~\bibnamefont{Aldegunde}}, \bibinfo {author}
  {\bibfnamefont{J.~M.}\ \bibnamefont{Hutson}},\ and\ \bibinfo {author}
  {\bibfnamefont{H.-C.}\ \bibnamefont{N\"agerl}},\ }%
  \bibfield{journal}{%
  \bibinfo {journal} {Nature Phys.}\ }%
  \textbf{\bibinfo {volume} {6}},\ \bibinfo {pages} {265} (\bibinfo {year}
  {2010})%
  \bibAnnoteFile{NoStop}{danzl2010}%
\bibitem{shuman2010}%
  \BibitemOpen
  \bibfield{author}{%
  \bibinfo {author} {\bibfnamefont{E.~S.}\ \bibnamefont{Shuman}}, \bibinfo
  {author} {\bibfnamefont{J.~F.}\ \bibnamefont{Barry}},\ and\ \bibinfo {author}
  {\bibfnamefont{D.}~\bibnamefont{DeMille}},\ }%
  \bibfield{journal}{%
  \bibinfo {journal} {Nature}\ }%
  \textbf{\bibinfo {volume} {467}},\ \bibinfo {pages} {820} (\bibinfo {year}
  {2010})%
  \bibAnnoteFile{NoStop}{shuman2010}%
\bibitem{zahzam2006}%
  \BibitemOpen
  \bibfield{author}{%
  \bibinfo {author} {\bibfnamefont{N.}~\bibnamefont{Zahzam}}, \bibinfo {author}
  {\bibfnamefont{T.}~\bibnamefont{Vogt}}, \bibinfo {author}
  {\bibfnamefont{M.}~\bibnamefont{Mudrich}}, \bibinfo {author}
  {\bibfnamefont{D.}~\bibnamefont{Comparat}},\ and\ \bibinfo {author}
  {\bibfnamefont{P.}~\bibnamefont{Pillet}},\ }%
  \bibfield{journal}{%
  \bibinfo {journal} {Phys. Rev. Lett.}\ }%
  \textbf{\bibinfo {volume} {96}},\ \bibinfo {pages} {023202} (\bibinfo {year}
  {2006})%
  \bibAnnoteFile{NoStop}{zahzam2006}%
\bibitem{staanum2006}%
  \BibitemOpen
  \bibfield{author}{%
  \bibinfo {author} {\bibfnamefont{P.}~\bibnamefont{Staanum}}, \bibinfo
  {author} {\bibfnamefont{S.~D.}\ \bibnamefont{Kraft}}, \bibinfo {author}
  {\bibfnamefont{J.}~\bibnamefont{Lange}}, \bibinfo {author}
  {\bibfnamefont{R.}~\bibnamefont{Wester}},\ and\ \bibinfo {author}
  {\bibfnamefont{M.}~\bibnamefont{Weidem\"{u}ller}},\ }%
  \bibfield{journal}{%
  \bibinfo {journal} {Phys. Rev. Lett.}\ }%
  \textbf{\bibinfo {volume} {96}},\ \bibinfo {pages} {023201} (\bibinfo {year}
  {2006})%
  \bibAnnoteFile{NoStop}{staanum2006}%
\bibitem{hudson2008}%
  \BibitemOpen
  \bibfield{author}{%
  \bibinfo {author} {\bibfnamefont{E.~R.}\ \bibnamefont{Hudson}}, \bibinfo
  {author} {\bibfnamefont{N.~B.}\ \bibnamefont{Gilfoy}}, \bibinfo {author}
  {\bibfnamefont{S.}~\bibnamefont{Kotochigova}}, \bibinfo {author}
  {\bibfnamefont{J.~M.}\ \bibnamefont{Sage}},\ and\ \bibinfo {author}
  {\bibfnamefont{D.}~\bibnamefont{DeMille}},\ }%
  \bibfield{journal}{%
  \bibinfo {journal} {Phys. Rev. Lett.}\ }%
  \textbf{\bibinfo {volume} {100}},\ \bibinfo {pages} {203201} (\bibinfo {year}
  {2008})%
  \bibAnnoteFile{NoStop}{hudson2008}%
\bibitem{hummon2010}%
  \BibitemOpen
  \bibfield{author}{%
  \bibinfo {author} {\bibfnamefont{M.~T.}\ \bibnamefont{Hummon}}, \bibinfo
  {author} {\bibfnamefont{T.~V.}\ \bibnamefont{Tscherbul}}, \bibinfo {author}
  {\bibfnamefont{J.}~\bibnamefont{Klos}}, \bibinfo {author}
  {\bibfnamefont{H.-I.}\ \bibnamefont{Lu}}, \bibinfo {author}
  {\bibfnamefont{E.}~\bibnamefont{Tsikata}}, \bibinfo {author}
  {\bibfnamefont{W.~C.}\ \bibnamefont{Campbell}}, \bibinfo {author}
  {\bibfnamefont{A.}~\bibnamefont{Dalgarno}},\ and\ \bibinfo {author}
  {\bibfnamefont{J.~M.}\ \bibnamefont{Doyle}},\ }%
  \bibfield{journal}{%
  \bibinfo {journal} {arXiv:1009.2513v1 [physics.atom-ph]}}%
   (\bibinfo {year} {2010})%
  \bibAnnoteFile{NoStop}{hummon2010}%
\bibitem{cvitas2005}%
  \BibitemOpen
  \bibfield{author}{%
  \bibinfo {author} {\bibfnamefont{M.~T.}\ \bibnamefont{Cvita\v{s}}}, \bibinfo
  {author} {\bibfnamefont{P.}~\bibnamefont{Sold\'an}}, \bibinfo {author}
  {\bibfnamefont{J.~M.}\ \bibnamefont{Hutson}}, \bibinfo {author}
  {\bibfnamefont{P.}~\bibnamefont{Honvault}},\ and\ \bibinfo {author}
  {\bibfnamefont{J.-M.}\ \bibnamefont{Launay}},\ }%
  \bibfield{journal}{%
  \bibinfo {journal} {Phys. Rev. Lett.}\ }%
  \textbf{\bibinfo {volume} {94}},\ \bibinfo {pages} {200402} (\bibinfo {year}
  {2005})%
  \bibAnnoteFile{NoStop}{cvitas2005}%
\bibitem{quemener2008}%
  \BibitemOpen
  \bibfield{author}{%
  \bibinfo {author} {\bibfnamefont{G.}~\bibnamefont{Qu\'em\'ener}}, \bibinfo
  {author} {\bibfnamefont{N.}~\bibnamefont{Balakrishnan}},\ and\ \bibinfo
  {author} {\bibfnamefont{R.~V.}\ \bibnamefont{Krems}},\ }%
  \bibfield{journal}{%
  \bibinfo {journal} {Phys. Rev. A}\ }%
  \textbf{\bibinfo {volume} {77}},\ \bibinfo {pages} {030704} (\bibinfo {year}
  {2008})%
  \bibAnnoteFile{NoStop}{quemener2008}%
\bibitem{krems2008}%
  \BibitemOpen
  \bibfield{author}{%
  \bibinfo {author} {\bibfnamefont{R.~V.}\ \bibnamefont{Krems}},\ }%
  \bibfield{journal}{%
  \bibinfo {journal} {Phys. Chem. Chem. Phys.}\ }%
  \textbf{\bibinfo {volume} {10}},\ \bibinfo {pages} {4079} (\bibinfo {year}
  {2008})%
  \bibAnnoteFile{NoStop}{krems2008}%
\bibitem{bell2009}%
  \BibitemOpen
  \bibfield{author}{%
  \bibinfo {author} {\bibfnamefont{M.}~\bibnamefont{Bell}}\ and\ \bibinfo
  {author} {\bibfnamefont{T.~P.}\ \bibnamefont{Softley}},\ }%
  \bibfield{journal}{%
  \bibinfo {journal} {Mol. Phys.}\ }%
  \textbf{\bibinfo {volume} {107}},\ \bibinfo {pages} {99} (\bibinfo {year}
  {20099})%
  \bibAnnoteFile{NoStop}{bell2009}%
\bibitem{kirste2010}%
  \BibitemOpen
  \bibfield{author}{%
  \bibinfo {author} {\bibfnamefont{M.}~\bibnamefont{Kirste}}, \bibinfo {author}
  {\bibfnamefont{L.}~\bibnamefont{Scharfenberg}}, \bibinfo {author}
  {\bibfnamefont{J.}~\bibnamefont{K\l{}os}}, \bibinfo {author}
  {\bibfnamefont{F.}~\bibnamefont{Lique}}, \bibinfo {author}
  {\bibfnamefont{M.~H.}\ \bibnamefont{Alexander}}, \bibinfo {author}
  {\bibfnamefont{G.}~\bibnamefont{Meijer}},\ and\ \bibinfo {author}
  {\bibfnamefont{S.~Y.~T.}\ \bibnamefont{van~de Meerakker}},\ }%
  \bibfield{journal}{%
  \bibinfo {journal} {Phys. Rev. A}\ }%
  \textbf{\bibinfo {volume} {82}},\ \bibinfo {pages} {042717} (\bibinfo {year}
  {2010})%
  \bibAnnoteFile{NoStop}{kirste2010}%
\bibitem{ni2010}%
  \BibitemOpen
  \bibfield{author}{%
  \bibinfo {author} {\bibfnamefont{K.-K.}\ \bibnamefont{Ni}}, \bibinfo {author}
  {\bibfnamefont{S.}~\bibnamefont{Ospelkaus}}, \bibinfo {author}
  {\bibfnamefont{D.}~\bibnamefont{Wang}}, \bibinfo {author}
  {\bibfnamefont{G.}~\bibnamefont{Qu\'em\'ener}}, \bibinfo {author}
  {\bibfnamefont{B.}~\bibnamefont{Neyenhuis}}, \bibinfo {author}
  {\bibfnamefont{M.~H.~G.}\ \bibnamefont{de~Miranda}}, \bibinfo {author}
  {\bibfnamefont{J.~L.}\ \bibnamefont{Bohn}}, \bibinfo {author}
  {\bibfnamefont{J.}~\bibnamefont{Ye}},\ and\ \bibinfo {author}
  {\bibfnamefont{D.~S.}\ \bibnamefont{Jin}},\ }%
  \bibfield{journal}{%
  \bibinfo {journal} {Nature}\ }%
  \textbf{\bibinfo {volume} {464}},\ \bibinfo {pages} {1324} (\bibinfo {year}
  {2010})%
  \bibAnnoteFile{NoStop}{ni2010}%
\bibitem{ospelkaus2010a}%
  \BibitemOpen
  \bibfield{author}{%
  \bibinfo {author} {\bibfnamefont{S.}~\bibnamefont{Ospelkaus}}, \bibinfo
  {author} {\bibfnamefont{K.-K.}\ \bibnamefont{Ni}}, \bibinfo {author}
  {\bibfnamefont{D.}~\bibnamefont{Wang}}, \bibinfo {author}
  {\bibfnamefont{M.~H.~G.}\ \bibnamefont{de~Miranda}}, \bibinfo {author}
  {\bibfnamefont{B.}~\bibnamefont{Neyenhuis}}, \bibinfo {author}
  {\bibfnamefont{G.}~\bibnamefont{Qu\'em\'ener}}, \bibinfo {author}
  {\bibfnamefont{P.~S.}\ \bibnamefont{Julienne}}, \bibinfo {author}
  {\bibfnamefont{J.}~\bibnamefont{Bohn}}, \bibinfo {author}
  {\bibfnamefont{D.~S.}\ \bibnamefont{Jin}},\ and\ \bibinfo {author}
  {\bibfnamefont{J.}~\bibnamefont{Ye}},\ }%
  \bibfield{journal}{%
  \bibinfo {journal} {Science}\ }%
  \textbf{\bibinfo {volume} {327}},\ \bibinfo {pages} {853} (\bibinfo {year}
  {2010})%
  \bibAnnoteFile{NoStop}{ospelkaus2010a}%
\bibitem{efimov1970}%
  \BibitemOpen
  \bibfield{author}{%
  \bibinfo {author} {\bibfnamefont{V.}~\bibnamefont{Efimov}},\ }%
  \bibfield{journal}{%
  \bibinfo {journal} {Phys. Lett. B}\ }%
  \textbf{\bibinfo {volume} {33}},\ \bibinfo {pages} {563} (\bibinfo {year}
  {1970})%
  \bibAnnoteFile{NoStop}{efimov1970}%
\bibitem{efimov2010}%
  \BibitemOpen
  \bibfield{author}{%
  \bibinfo {author} {\bibfnamefont{V.}~\bibnamefont{Efimov}},\ }%
  \bibfield{journal}{%
  \bibinfo {journal} {Nature Phys.}\ }%
  \textbf{\bibinfo {volume} {5}},\ \bibinfo {pages} {533} (\bibinfo {year}
  {2010})%
  \bibAnnoteFile{NoStop}{efimov2010}%
\bibitem{micheli2006}%
  \BibitemOpen
  \bibfield{author}{%
  \bibinfo {author} {\bibfnamefont{A.}~\bibnamefont{Micheli}}, \bibinfo
  {author} {\bibfnamefont{G.~K.}\ \bibnamefont{Brennen}},\ and\ \bibinfo
  {author} {\bibfnamefont{P.}~\bibnamefont{Zoller}},\ }%
  \bibfield{journal}{%
  \bibinfo {journal} {Nature Physics}\ }%
  \textbf{\bibinfo {volume} {2}},\ \bibinfo {pages} {341} (\bibinfo {year}
  {2006})%
  \bibAnnoteFile{NoStop}{micheli2006}%
\bibitem{demille2002}%
  \BibitemOpen
  \bibfield{author}{%
  \bibinfo {author} {\bibfnamefont{D.}~\bibnamefont{DeMille}},\ }%
  \bibfield{journal}{%
  \bibinfo {journal} {Phys. Rev. Lett.}\ }%
  \textbf{\bibinfo {volume} {88}},\ \bibinfo {pages} {067901} (\bibinfo {year}
  {2002})%
  \bibAnnoteFile{NoStop}{demille2002}%
\bibitem{rabl2006}%
  \BibitemOpen
  \bibfield{author}{%
  \bibinfo {author} {\bibfnamefont{P.}~\bibnamefont{Rabl}}, \bibinfo {author}
  {\bibfnamefont{D.}~\bibnamefont{DeMille}}, \bibinfo {author}
  {\bibfnamefont{J.~M.}\ \bibnamefont{Doyle}}, \bibinfo {author}
  {\bibfnamefont{M.~D.}\ \bibnamefont{Lukin}}, \bibinfo {author}
  {\bibfnamefont{R.~J.}\ \bibnamefont{Schoelkopf}},\ and\ \bibinfo {author}
  {\bibfnamefont{P.}~\bibnamefont{Zoller}},\ }%
  \bibfield{journal}{%
  \bibinfo {journal} {Phys. Rev. Lett.}\ }%
  \textbf{\bibinfo {volume} {97}},\ \bibinfo {pages} {033003} (\bibinfo {year}
  {2006})%
  \bibAnnoteFile{NoStop}{rabl2006}%
\bibitem{demille2008}%
  \BibitemOpen
  \bibfield{author}{%
  \bibinfo {author} {\bibfnamefont{D.}~\bibnamefont{DeMille}}, \bibinfo
  {author} {\bibfnamefont{S.}~\bibnamefont{Sainis}}, \bibinfo {author}
  {\bibfnamefont{J.}~\bibnamefont{Sage}}, \bibinfo {author}
  {\bibfnamefont{T.}~\bibnamefont{Bergeman}}, \bibinfo {author}
  {\bibfnamefont{S.}~\bibnamefont{Kotochigova}},\ and\ \bibinfo {author}
  {\bibfnamefont{E.}~\bibnamefont{Tiesinga}},\ }%
  \bibfield{journal}{%
  \bibinfo {journal} {Phys. Rev. Lett.}\ }%
  \textbf{\bibinfo {volume} {100}},\ \bibinfo {pages} {043202} (\bibinfo {year}
  {2008})%
  \bibAnnoteFile{NoStop}{demille2008}%
\bibitem{kajita2008}%
  \BibitemOpen
  \bibfield{author}{%
  \bibinfo {author} {\bibfnamefont{M.}~\bibnamefont{Kajita}},\ }%
  \bibfield{journal}{%
  \bibinfo {journal} {Phys. Rev. A}\ }%
  \textbf{\bibinfo {volume} {77}},\ \bibinfo {pages} {012511} (\bibinfo {year}
  {2008})%
  \bibAnnoteFile{NoStop}{kajita2008}%
\bibitem{zelevinsky2008}%
  \BibitemOpen
  \bibfield{author}{%
  \bibinfo {author} {\bibfnamefont{T.}~\bibnamefont{Zelevinsky}}, \bibinfo
  {author} {\bibfnamefont{S.}~\bibnamefont{Kotochigova}},\ and\ \bibinfo
  {author} {\bibfnamefont{J.}~\bibnamefont{Ye}},\ }%
  \bibfield{journal}{%
  \bibinfo {journal} {Phys. Rev. Lett.}\ }%
  \textbf{\bibinfo {volume} {100}},\ \bibinfo {pages} {043201} (\bibinfo {year}
  {2008})%
  \bibAnnoteFile{NoStop}{zelevinsky2008}%
\bibitem{tarbutt2009}%
  \BibitemOpen
  \bibfield{author}{%
  \bibinfo {author} {\bibfnamefont{M.~R.}\ \bibnamefont{Tarbutt}}, \bibinfo
  {author} {\bibfnamefont{J.~J.}\ \bibnamefont{Hudson}}, \bibinfo {author}
  {\bibfnamefont{B.~E.}\ \bibnamefont{Sauer}},\ and\ \bibinfo {author}
  {\bibfnamefont{E.~A.}\ \bibnamefont{Hinds}},\ }%
  \bibfield{journal}{%
  \bibinfo {journal} {arXiv}\ }%
  \textbf{\bibinfo {volume} {[physics.atom-phys]}},\ \bibinfo {pages}
  {0811.2950v1} (\bibinfo {year} {2009})%
  \bibAnnoteFile{NoStop}{tarbutt2009}%
\bibitem{klemperer2006}%
  \BibitemOpen
  \bibfield{author}{%
  \bibinfo {author} {\bibfnamefont{W.}~\bibnamefont{Klemperer}},\ }%
  \bibfield{journal}{%
  \bibinfo {journal} {Proc. Nat. Acad. Sci.}\ }%
  \textbf{\bibinfo {volume} {103}},\ \bibinfo {pages} {12232} (\bibinfo {year}
  {2006})%
  \bibAnnoteFile{NoStop}{klemperer2006}%
\bibitem{herbst2008}%
  \BibitemOpen
  \bibfield{author}{%
  \bibinfo {author} {\bibfnamefont{E.}~\bibnamefont{Herbst}}\ and\ \bibinfo
  {author} {\bibfnamefont{T.~J.}\ \bibnamefont{Millar}},\ }%
  \enquote{\bibinfo {title} {The chemistry of cold interstellar cloud cores},}\
  in\ \emph{\bibinfo {booktitle} {Low temperatures and cold molecules}},\
  \bibinfo {editor} {edited by\ \bibinfo {editor} {\bibfnamefont{I.~W.~M.}\
  \bibnamefont{Smith}}}\ (\bibinfo {publisher} {World Scientific Publishing},\
  \bibinfo {year} {2008})\ p.~\bibinfo {pages} {1}%
  \bibAnnoteFile{NoStop}{herbst2008}%
\bibitem{gerlich2008b}%
  \BibitemOpen
  \bibfield{author}{%
  \bibinfo {author} {\bibfnamefont{D.}~\bibnamefont{Gerlich}},\ }%
  \enquote{\bibinfo {title} {The production and study of ultra-cold molecular
  ions},}\ in\ \emph{\bibinfo {booktitle} {Low temperatures and cold
  molecules}},\ \bibinfo {editor} {edited by\ \bibinfo {editor}
  {\bibfnamefont{I.~W.~M.}\ \bibnamefont{Smith}}}\ (\bibinfo {publisher} {World
  Scientific Publishing},\ \bibinfo {year} {2008})\ p.\ \bibinfo {pages} {295}%
  \bibAnnoteFile{NoStop}{gerlich2008b}%
\bibitem{willitsch2008}%
  \BibitemOpen
  \bibfield{author}{%
  \bibinfo {author} {\bibfnamefont{S.}~\bibnamefont{Willitsch}}, \bibinfo
  {author} {\bibfnamefont{M.~T.}\ \bibnamefont{Bell}}, \bibinfo {author}
  {\bibfnamefont{A.~D.}\ \bibnamefont{Gingell}}, \bibinfo {author}
  {\bibfnamefont{S.~R.}\ \bibnamefont{Procter}},\ and\ \bibinfo {author}
  {\bibfnamefont{T.~P.}\ \bibnamefont{Softley}},\ }%
  \bibfield{journal}{%
  \bibinfo {journal} {Phys. Rev. Lett.}\ }%
  \textbf{\bibinfo {volume} {100}},\ \bibinfo {pages} {043203} (\bibinfo {year}
  {2008})%
  \bibAnnoteFile{NoStop}{willitsch2008}%
\bibitem{schneider2010}%
  \BibitemOpen
  \bibfield{author}{%
  \bibinfo {author} {\bibfnamefont{T.}~\bibnamefont{Schneider}}, \bibinfo
  {author} {\bibfnamefont{B.}~\bibnamefont{Roth}}, \bibinfo {author}
  {\bibfnamefont{H.}~\bibnamefont{Duncker}}, \bibinfo {author}
  {\bibfnamefont{I.}~\bibnamefont{Ernsting}},\ and\ \bibinfo {author}
  {\bibfnamefont{S.}~\bibnamefont{Schiller}},\ }%
  \bibfield{journal}{%
  \bibinfo {journal} {Nature Phys.}\ }%
  \textbf{\bibinfo {volume} {6}},\ \bibinfo {pages} {275} (\bibinfo {year}
  {2010})%
  \bibAnnoteFile{NoStop}{schneider2010}%
\bibitem{staanum2010}%
  \BibitemOpen
  \bibfield{author}{%
  \bibinfo {author} {\bibfnamefont{P.~F.}\ \bibnamefont{Staanum}}, \bibinfo
  {author} {\bibfnamefont{K.}~\bibnamefont{Hojbjerre}}, \bibinfo {author}
  {\bibfnamefont{P.~S.}\ \bibnamefont{Skyt}}, \bibinfo {author}
  {\bibfnamefont{A.~K.}\ \bibnamefont{Hansen}},\ and\ \bibinfo {author}
  {\bibfnamefont{M.}~\bibnamefont{Drewsen}},\ }%
  \bibfield{journal}{%
  \bibinfo {journal} {Nature Physics}\ }%
  \textbf{\bibinfo {volume} {6}},\ \bibinfo {pages} {271} (\bibinfo {year}
  {2010})%
  \bibAnnoteFile{NoStop}{staanum2010}%
\bibitem{tong2010}%
  \BibitemOpen
  \bibfield{author}{%
  \bibinfo {author} {\bibfnamefont{X.}~\bibnamefont{Tong}}, \bibinfo {author}
  {\bibfnamefont{A.~H.}\ \bibnamefont{Winney}},\ and\ \bibinfo {author}
  {\bibfnamefont{S.}~\bibnamefont{Willitsch}},\ }%
  \bibfield{journal}{%
  \bibinfo {journal} {Phys. Rev. Lett.}\ }%
  \textbf{\bibinfo {volume} {105}},\ \bibinfo {pages} {143001} (\bibinfo {year}
  {2010})%
  \bibAnnoteFile{NoStop}{tong2010}%
\bibitem{schiller2005}%
  \BibitemOpen
  \bibfield{author}{%
  \bibinfo {author} {\bibfnamefont{S.}~\bibnamefont{Schiller}}\ and\ \bibinfo
  {author} {\bibfnamefont{V.}~\bibnamefont{Korobov}},\ }%
  \bibfield{journal}{%
  \bibinfo {journal} {Phys. Rev. A}\ }%
  \textbf{\bibinfo {volume} {71}},\ \bibinfo {pages} {032505} (\bibinfo {year}
  {2005})%
  \bibAnnoteFile{NoStop}{schiller2005}%
\bibitem{meyer2006}%
  \BibitemOpen
  \bibfield{author}{%
  \bibinfo {author} {\bibfnamefont{E.~R.}\ \bibnamefont{Meyer}}, \bibinfo
  {author} {\bibfnamefont{J.~L.}\ \bibnamefont{Bohn}},\ and\ \bibinfo {author}
  {\bibfnamefont{M.~P.}\ \bibnamefont{Deskevich}},\ }%
  \bibfield{journal}{%
  \bibinfo {journal} {Phys. Rev. A}\ }%
  \textbf{\bibinfo {volume} {73}},\ \bibinfo {pages} {062108} (\bibinfo {year}
  {2006})%
  \bibAnnoteFile{NoStop}{meyer2006}%
\bibitem{meyer2009}%
  \BibitemOpen
  \bibfield{author}{%
  \bibinfo {author} {\bibfnamefont{E.~R.}\ \bibnamefont{Meyer}}\ and\ \bibinfo
  {author} {\bibfnamefont{J.~L.}\ \bibnamefont{Bohn}},\ }%
  \bibfield{journal}{%
  \bibinfo {journal} {Phys. Rev. A}\ }%
  \textbf{\bibinfo {volume} {80}},\ \bibinfo {pages} {042508} (\bibinfo {year}
  {2009})%
  \bibAnnoteFile{NoStop}{meyer2009}%
\bibitem{molhave2000}%
  \BibitemOpen
  \bibfield{author}{%
  \bibinfo {author} {\bibfnamefont{K.}~\bibnamefont{M\o{lhave}}}\ and\ \bibinfo
  {author} {\bibfnamefont{M.}~\bibnamefont{Drewsen}},\ }%
  \bibfield{journal}{%
  \bibinfo {journal} {Phys. Rev. A}\ }%
  \textbf{\bibinfo {volume} {62}},\ \bibinfo {pages} {011401(R)} (\bibinfo
  {year} {2000})%
  \bibAnnoteFile{NoStop}{molhave2000}%
\bibitem{grier2009}%
  \BibitemOpen
  \bibfield{author}{%
  \bibinfo {author} {\bibfnamefont{A.~T.}\ \bibnamefont{Grier}}, \bibinfo
  {author} {\bibfnamefont{M.}~\bibnamefont{Cetina}}, \bibinfo {author}
  {\bibfnamefont{F.}~\bibnamefont{Oru\ifmmode \check{c}\else
  \v{c}\fi{}evi\ifmmode~\acute{c}\else \'{c}\fi{}}},\ and\ \bibinfo {author}
  {\bibfnamefont{V.}~\bibnamefont{Vuleti\ifmmode~\acute{c}\else \'{c}\fi{}}},\
  }%
  \bibfield{journal}{%
  \bibinfo {journal} {Phys. Rev. Lett.}\ }%
  \textbf{\bibinfo {volume} {102}},\ \bibinfo {pages} {223201} (\bibinfo {year}
  {2009})%
  \bibAnnoteFile{NoStop}{grier2009}%
\bibitem{zipkes2010}%
  \BibitemOpen
  \bibfield{author}{%
  \bibinfo {author} {\bibfnamefont{C.}~\bibnamefont{Zipkes}}, \bibinfo {author}
  {\bibfnamefont{S.}~\bibnamefont{Palzer}}, \bibinfo {author}
  {\bibfnamefont{L.}~\bibnamefont{Ratschbacher}}, \bibinfo {author}
  {\bibfnamefont{C.}~\bibnamefont{Sias}},\ and\ \bibinfo {author}
  {\bibfnamefont{M.}~\bibnamefont{K\"ohl}},\ }%
  \bibfield{journal}{%
  \bibinfo {journal} {Nature}\ }%
  \textbf{\bibinfo {volume} {464}},\ \bibinfo {pages} {308} (\bibinfo {year}
  {2010})%
  \bibAnnoteFile{NoStop}{zipkes2010}%
\bibitem{zipkes2010a}%
  \BibitemOpen
  \bibfield{author}{%
  \bibinfo {author} {\bibfnamefont{C.}~\bibnamefont{Zipkes}}, \bibinfo {author}
  {\bibfnamefont{S.}~\bibnamefont{Palzer}}, \bibinfo {author}
  {\bibfnamefont{L.}~\bibnamefont{Ratschbacher}}, \bibinfo {author}
  {\bibfnamefont{C.}~\bibnamefont{Sias}},\ and\ \bibinfo {author}
  {\bibfnamefont{M.}~\bibnamefont{K\"ohl}},\ }%
  \bibfield{journal}{%
  \bibinfo {journal} {Phys. Rev. Lett.}\ }%
  \textbf{\bibinfo {volume} {105}},\ \bibinfo {pages} {133201} (\bibinfo {year}
  {2010})%
  \bibAnnoteFile{NoStop}{zipkes2010a}%
\bibitem{schmid2010}%
  \BibitemOpen
  \bibfield{author}{%
  \bibinfo {author} {\bibfnamefont{S.}~\bibnamefont{Schmid}}, \bibinfo {author}
  {\bibfnamefont{A.}~\bibnamefont{H\"arter}},\ and\ \bibinfo {author}
  {\bibfnamefont{J.~H.}\ \bibnamefont{Denschlag}},\ }%
  \bibfield{journal}{%
  \bibinfo {journal} {Phys. Rev. Lett.}\ }%
  \textbf{\bibinfo {volume} {105}},\ \bibinfo {pages} {133202} (\bibinfo {year}
  {2010})%
  \bibAnnoteFile{NoStop}{schmid2010}%
\bibitem{idziaszek2009}%
  \BibitemOpen
  \bibfield{author}{%
  \bibinfo {author} {\bibfnamefont{Z.}~\bibnamefont{Idziaszek}}, \bibinfo
  {author} {\bibfnamefont{T.}~\bibnamefont{Calarco}}, \bibinfo {author}
  {\bibfnamefont{P.~S.}\ \bibnamefont{Julienne}},\ and\ \bibinfo {author}
  {\bibfnamefont{A.}~\bibnamefont{Simoni}},\ }%
  \bibfield{journal}{%
  \bibinfo {journal} {Phys. Rev. A}\ }%
  \textbf{\bibinfo {volume} {79}},\ \bibinfo {pages} {010702} (\bibinfo {year}
  {2009})%
  \bibAnnoteFile{NoStop}{idziaszek2009}%
\bibitem{jones2006}%
  \BibitemOpen
  \bibfield{author}{%
  \bibinfo {author} {\bibfnamefont{K.~M.}\ \bibnamefont{Jones}}, \bibinfo
  {author} {\bibfnamefont{E.}~\bibnamefont{Tiesinga}}, \bibinfo {author}
  {\bibfnamefont{P.~D.}\ \bibnamefont{Lett}},\ and\ \bibinfo {author}
  {\bibfnamefont{P.~S.}\ \bibnamefont{Julienne}},\ }%
  \bibfield{journal}{%
  \bibinfo {journal} {Rev. Mod. Phys.}\ }%
  \textbf{\bibinfo {volume} {78}},\ \bibinfo {pages} {483} (\bibinfo {year}
  {2006})%
  \bibAnnoteFile{NoStop}{jones2006}%
\bibitem{chin2010}%
  \BibitemOpen
  \bibfield{author}{%
  \bibinfo {author} {\bibfnamefont{C.}~\bibnamefont{Chin}}, \bibinfo {author}
  {\bibfnamefont{R.}~\bibnamefont{Grimm}}, \bibinfo {author}
  {\bibfnamefont{P.}~\bibnamefont{Julienne}},\ and\ \bibinfo {author}
  {\bibfnamefont{E.}~\bibnamefont{Tiesinga}},\ }%
  \bibfield{journal}{%
  \bibinfo {journal} {Rev. Mod. Phys.}\ }%
  \textbf{\bibinfo {volume} {82}},\ \bibinfo {pages} {1225} (\bibinfo {year}
  {2010})%
  \bibAnnoteFile{NoStop}{chin2010}%
\bibitem{makarov2003}%
  \BibitemOpen
  \bibfield{author}{%
  \bibinfo {author} {\bibfnamefont{O.~P.}\ \bibnamefont{Makarov}}, \bibinfo
  {author} {\bibfnamefont{R.}~\bibnamefont{C\^ot\'e}}, \bibinfo {author}
  {\bibfnamefont{H.~M.~W.}\ \bibnamefont{W.}},\ and\ \bibinfo {author}
  {\bibnamefont{Smith}},\ }%
  \bibfield{journal}{%
  \bibinfo {journal} {Phys. Rev. A}\ }%
  \textbf{\bibinfo {volume} {67}},\ \bibinfo {pages} {042705} (\bibinfo {year}
  {2003})%
  \bibAnnoteFile{NoStop}{makarov2003}%
\bibitem{guerout2010}%
  \BibitemOpen
  \bibfield{author}{%
  \bibinfo {author} {\bibfnamefont{R.}~\bibnamefont{Gu\'erout}}, \bibinfo
  {author} {\bibfnamefont{M.}~\bibnamefont{Aymar}},\ and\ \bibinfo {author}
  {\bibfnamefont{O.}~\bibnamefont{Dulieu}},\ }%
  \bibfield{journal}{%
  \bibinfo {journal} {Phys. Rev. A}\ }%
  \textbf{\bibinfo {volume} {82}},\ \bibinfo {pages} {042508} (\bibinfo {year}
  {2010})%
  \bibAnnoteFile{NoStop}{guerout2010}%
\bibitem{removille2009}%
  \BibitemOpen
  \bibfield{author}{%
  \bibinfo {author} {\bibfnamefont{S.}~\bibnamefont{Removille}}, \bibinfo
  {author} {\bibfnamefont{R.}~\bibnamefont{Dubessy}}, \bibinfo {author}
  {\bibfnamefont{B.}~\bibnamefont{Dubost}}, \bibinfo {author}
  {\bibfnamefont{Q.}~\bibnamefont{Glorieux}}, \bibinfo {author}
  {\bibfnamefont{T.}~\bibnamefont{Coudreau}}, \bibinfo {author}
  {\bibfnamefont{S.}~\bibnamefont{Guibal}}, \bibinfo {author}
  {\bibfnamefont{J.-P.}\ \bibnamefont{Likforman}},\ and\ \bibinfo {author}
  {\bibfnamefont{L.}~\bibnamefont{Guidoni}},\ }%
  \bibfield{journal}{%
  \bibinfo {journal} {J. Phys. B: At. Molec. Opt. Phys.}\ }%
  \textbf{\bibinfo {volume} {42}},\ \bibinfo {pages} {154014} (\bibinfo {year}
  {2009})%
  \bibAnnoteFile{NoStop}{removille2009}%
\bibitem{removille2009a}%
  \BibitemOpen
  \bibfield{author}{%
  \bibinfo {author} {\bibfnamefont{S.}~\bibnamefont{Removille}}, \bibinfo
  {author} {\bibfnamefont{R.}~\bibnamefont{Dubessy}}, \bibinfo {author}
  {\bibfnamefont{Q.}~\bibnamefont{Glorieux}}, \bibinfo {author}
  {\bibfnamefont{S.}~\bibnamefont{Guibal}}, \bibinfo {author}
  {\bibfnamefont{T.}~\bibnamefont{Coudreau}}, \bibinfo {author}
  {\bibfnamefont{L.}~\bibnamefont{Guidoni}},\ and\ \bibinfo {author}
  {\bibfnamefont{J.-P.}\ \bibnamefont{Likforman}},\ }%
  \bibfield{journal}{%
  \bibinfo {journal} {App. Phys. B}\ }%
  \textbf{\bibinfo {volume} {97}},\ \bibinfo {pages} {47} (\bibinfo {year}
  {2009})%
  \bibAnnoteFile{NoStop}{removille2009a}%
\bibitem{aymar2005}%
  \BibitemOpen
  \bibfield{author}{%
  \bibinfo {author} {\bibfnamefont{M.}~\bibnamefont{Aymar}}\ and\ \bibinfo
  {author} {\bibfnamefont{O.}~\bibnamefont{Dulieu}},\ }%
  \bibfield{journal}{%
  \bibinfo {journal} {J. Chem. Phys.}\ }%
  \textbf{\bibinfo {volume} {122}},\ \bibinfo {pages} {204302} (\bibinfo {year}
  {2005})%
  \bibAnnoteFile{NoStop}{aymar2005}%
\bibitem{durand1974}%
  \BibitemOpen
  \bibfield{author}{%
  \bibinfo {author} {\bibfnamefont{P.}~\bibnamefont{Durand}}\ and\ \bibinfo
  {author} {\bibfnamefont{J.}~\bibnamefont{Barthelat}},\ }%
  \bibfield{journal}{%
  \bibinfo {journal} {Chem. Phys. Lett.}\ }%
  \textbf{\bibinfo {volume} {27}},\ \bibinfo {pages} {191} (\bibinfo {year}
  {1974})%
  \bibAnnoteFile{NoStop}{durand1974}%
\bibitem{durand1975}%
  \BibitemOpen
  \bibfield{author}{%
  \bibinfo {author} {\bibfnamefont{P.}~\bibnamefont{Durand}}\ and\ \bibinfo
  {author} {\bibfnamefont{J.}~\bibnamefont{Barthelat}},\ }%
  \bibfield{journal}{%
  \bibinfo {journal} {Theor. chim. Acta}\ }%
  \textbf{\bibinfo {volume} {38}},\ \bibinfo {pages} {283} (\bibinfo {year}
  {1975})%
  \bibAnnoteFile{NoStop}{durand1975}%
\bibitem{fuentealba1985}%
  \BibitemOpen
  \bibfield{author}{%
  \bibinfo {author} {\bibfnamefont{P.}~\bibnamefont{Fuentealba}}, \bibinfo
  {author} {\bibfnamefont{L.}~\bibnamefont{von Szentpaly}}, \bibinfo {author}
  {\bibfnamefont{H.}~\bibnamefont{Preuss}},\ and\ \bibinfo {author}
  {\bibfnamefont{H.}~\bibnamefont{Stoll}},\ }%
  \bibfield{journal}{%
  \bibinfo {journal} {J. Phys. B: At. and Mol. Phys.}\ }%
  \textbf{\bibinfo {volume} {18}},\ \bibinfo {pages} {1287} (\bibinfo {year}
  {1985})%
  \bibAnnoteFile{NoStop}{fuentealba1985}%
\bibitem{fuentealba1987}%
  \BibitemOpen
  \bibfield{author}{%
  \bibinfo {author} {\bibfnamefont{P.}~\bibnamefont{Fuentealba}}\ and\ \bibinfo
  {author} {\bibfnamefont{O.}~\bibnamefont{Reyes}},\ }%
  \bibfield{journal}{%
  \bibinfo {journal} {Molec. Phys.}\ }%
  \textbf{\bibinfo {volume} {62}},\ \bibinfo {pages} {1291} (\bibinfo {year}
  {1987})%
  \bibAnnoteFile{NoStop}{fuentealba1987}%
\bibitem{muller1984}%
  \BibitemOpen
  \bibfield{author}{%
  \bibinfo {author} {\bibfnamefont{W.}~\bibnamefont{M\"uller}}\ and\ \bibinfo
  {author} {\bibfnamefont{W.}~\bibnamefont{Meyer}},\ }%
  \bibfield{journal}{%
  \bibinfo {journal} {J. Chem. Phys.}\ }%
  \textbf{\bibinfo {volume} {80}},\ \bibinfo {pages} {3311} (\bibinfo {year}
  {1984})%
  \bibAnnoteFile{NoStop}{muller1984}%
\bibitem{guerout2010a}%
  \BibitemOpen
  \bibfield{author}{%
  \bibinfo {author} {\bibfnamefont{R.}~\bibnamefont{Gu\'erout}}, \bibinfo
  {author} {\bibfnamefont{O.}~\bibnamefont{Dulieu}},\ and\ \bibinfo {author}
  {\bibfnamefont{F.}~\bibnamefont{Spiegelman}},\ }%
  \bibfield{journal}{%
  \bibinfo {journal} {in preparation}}%
   (\bibinfo {year} {2011})%
  \bibAnnoteFile{NoStop}{guerout2010a}%
\bibitem{huron1973}%
  \BibitemOpen
  \bibfield{author}{%
  \bibinfo {author} {\bibfnamefont{B.}~\bibnamefont{Huron}}, \bibinfo {author}
  {\bibfnamefont{J.-P.}\ \bibnamefont{Malrieu}},\ and\ \bibinfo {author}
  {\bibfnamefont{P.}~\bibnamefont{Rancurel}},\ }%
  \bibfield{journal}{%
  \bibinfo {journal} {J. Chem. Phys.}\ }%
  \textbf{\bibinfo {volume} {58}},\ \bibinfo {pages} {5745} (\bibinfo {year}
  {1973})%
  \bibAnnoteFile{NoStop}{huron1973}%
\bibitem{boutassetta1996}%
  \BibitemOpen
  \bibfield{author}{%
  \bibinfo {author} {\bibfnamefont{N.}~\bibnamefont{Boutassetta}}, \bibinfo
  {author} {\bibfnamefont{A.~R.}\ \bibnamefont{Allouche}},\ and\ \bibinfo
  {author} {\bibfnamefont{M.}~\bibnamefont{Aubert-Fr\'econ}},\ }%
  \bibfield{journal}{%
  \bibinfo {journal} {Phys. Rev. A}\ }%
  \textbf{\bibinfo {volume} {53}},\ \bibinfo {pages} {3845} (\bibinfo {year}
  {1996})%
  \bibAnnoteFile{NoStop}{boutassetta1996}%
\bibitem{coker1976}%
  \BibitemOpen
  \bibfield{author}{%
  \bibinfo {author} {\bibfnamefont{H.}~\bibnamefont{Coker}},\ }%
  \bibfield{journal}{%
  \bibinfo {journal} {J. Phys. Chem.}\ }%
  \textbf{\bibinfo {volume} {80}},\ \bibinfo {pages} {2078} (\bibinfo {year}
  {1976})%
  \bibAnnoteFile{NoStop}{coker1976}%
\bibitem{wilson1970}%
  \BibitemOpen
  \bibfield{author}{%
  \bibinfo {author} {\bibfnamefont{J.~N.}\ \bibnamefont{Wilson}}\ and\ \bibinfo
  {author} {\bibfnamefont{R.~M.}\ \bibnamefont{Curtis}},\ }%
  \bibfield{journal}{%
  \bibinfo {journal} {J. Chem. Phys.}\ }%
  \textbf{\bibinfo {volume} {74}},\ \bibinfo {pages} {187} (\bibinfo {year}
  {1970})%
  \bibAnnoteFile{NoStop}{wilson1970}%
\bibitem{bouissou2010}%
  \BibitemOpen
  \bibfield{author}{%
  \bibinfo {author} {\bibfnamefont{T.}~\bibnamefont{Bouissou}}, \bibinfo
  {author} {\bibfnamefont{G.}~\bibnamefont{Durand}}, \bibinfo {author}
  {\bibfnamefont{M.-C.}\ \bibnamefont{Heitz}},\ and\ \bibinfo {author}
  {\bibfnamefont{F.}~\bibnamefont{Spiegelman}},\ }%
  \bibfield{journal}{%
  \bibinfo {journal} {J. Chem. Phys.}\ }%
  \textbf{\bibinfo {volume} {133}},\ \bibinfo {pages} {164317} (\bibinfo {year}
  {2010})%
  \bibAnnoteFile{NoStop}{bouissou2010}%
\bibitem{barklem2002}%
  \BibitemOpen
  \bibfield{author}{%
  \bibinfo {author} {\bibfnamefont{P.~S.}\ \bibnamefont{Barklem}}\ and\
  \bibinfo {author} {\bibfnamefont{P.~J.}\ \bibnamefont{O'Mara}},\ }%
  \bibfield{journal}{%
  \bibinfo {journal} {Month. Not. Roy. Astron. Soc.}\ }%
  \textbf{\bibinfo {volume} {311}},\ \bibinfo {pages} {535} (\bibinfo {year}
  {2002})%
  \bibAnnoteFile{NoStop}{barklem2002}%
\bibitem{patil1997}%
  \BibitemOpen
  \bibfield{author}{%
  \bibinfo {author} {\bibfnamefont{S.~H.}\ \bibnamefont{Patil}}\ and\ \bibinfo
  {author} {\bibfnamefont{K.~T.}\ \bibnamefont{Tang}},\ }%
  \bibfield{journal}{%
  \bibinfo {journal} {J. Chem. Phys.}\ }%
  \textbf{\bibinfo {volume} {106}},\ \bibinfo {pages} {2298} (\bibinfo {year}
  {1997})%
  \bibAnnoteFile{NoStop}{patil1997}%
\bibitem{mitroy2008a}%
  \BibitemOpen
  \bibfield{author}{%
  \bibinfo {author} {\bibfnamefont{J.}~\bibnamefont{Mitroy}}, \bibinfo {author}
  {\bibfnamefont{J.~Y.}\ \bibnamefont{Zhang}},\ and\ \bibinfo {author}
  {\bibfnamefont{M.~W.~J.}\ \bibnamefont{Bromley}},\ }%
  \bibfield{journal}{%
  \bibinfo {journal} {Phys. Rev. A}\ }%
  \textbf{\bibinfo {volume} {77}},\ \bibinfo {pages} {032512} (\bibinfo {year}
  {2008})%
  \bibAnnoteFile{NoStop}{mitroy2008a}%
\bibitem{margolis2003}%
  \BibitemOpen
  \bibfield{author}{%
  \bibinfo {author} {\bibfnamefont{H.~S.}\ \bibnamefont{Margolis}}, \bibinfo
  {author} {\bibfnamefont{G.}~\bibnamefont{Huang}}, \bibinfo {author}
  {\bibfnamefont{G.~P.}\ \bibnamefont{Barwood}}, \bibinfo {author}
  {\bibfnamefont{S.~N.}\ \bibnamefont{Lea}}, \bibinfo {author}
  {\bibfnamefont{H.~A.}\ \bibnamefont{Klein}}, \bibinfo {author}
  {\bibfnamefont{W.~R.~C.}\ \bibnamefont{Rowley}}, \bibinfo {author}
  {\bibfnamefont{P.}~\bibnamefont{Gill}},\ and\ \bibinfo {author}
  {\bibfnamefont{R.~S.}\ \bibnamefont{Windeler}},\ }%
  \bibfield{journal}{%
  \bibinfo {journal} {Phys. Rev. A}\ }%
  \textbf{\bibinfo {volume} {67}},\ \bibinfo {pages} {032501} (\bibinfo {year}
  {2003})%
  \bibAnnoteFile{NoStop}{margolis2003}%
\bibitem{madej2004}%
  \BibitemOpen
  \bibfield{author}{%
  \bibinfo {author} {\bibfnamefont{A.~A.}\ \bibnamefont{Madej}}, \bibinfo
  {author} {\bibfnamefont{J.~E.}\ \bibnamefont{Bernard}}, \bibinfo {author}
  {\bibfnamefont{P.}~\bibnamefont{Dub\'e}}, \bibinfo {author}
  {\bibfnamefont{L.}~\bibnamefont{Marmet}},\ and\ \bibinfo {author}
  {\bibfnamefont{R.~S.}\ \bibnamefont{Windeler}},\ }%
  \bibfield{journal}{%
  \bibinfo {journal} {Phys. Rev. A}\ }%
  \textbf{\bibinfo {volume} {70}},\ \bibinfo {pages} {012507} (\bibinfo {year}
  {2004})%
  \bibAnnoteFile{NoStop}{madej2004}%
\bibitem{jiang2009}%
  \BibitemOpen
  \bibfield{author}{%
  \bibinfo {author} {\bibfnamefont{D.}~\bibnamefont{Jiang}}, \bibinfo {author}
  {\bibfnamefont{B.}~\bibnamefont{Arora}}, \bibinfo {author}
  {\bibfnamefont{M.~S.}\ \bibnamefont{Safronova}},\ and\ \bibinfo {author}
  {\bibfnamefont{C.~W.}\ \bibnamefont{Clark}},\ }%
  \bibfield{journal}{%
  \bibinfo {journal} {J. Phys. B: At. Mol.Opt. Phys.}\ }%
  \textbf{\bibinfo {volume} {42}},\ \bibinfo {pages} {154020} (\bibinfo {year}
  {2009})%
  \bibAnnoteFile{NoStop}{jiang2009}%
\bibitem{cohen1965}%
  \BibitemOpen
  \bibfield{author}{%
  \bibinfo {author} {\bibfnamefont{H.~D.}\ \bibnamefont{Cohen}}\ and\ \bibinfo
  {author} {\bibfnamefont{C.~C.~J.}\ \bibnamefont{Roothaan}},\ }%
  \bibfield{journal}{%
  \bibinfo {journal} {J. Chem. Phys.}\ }%
  \textbf{\bibinfo {volume} {43}},\ \bibinfo {pages} {S34} (\bibinfo {year}
  {1965})%
  \bibAnnoteFile{NoStop}{cohen1965}%
\bibitem{sadlej1991}%
  \BibitemOpen
  \bibfield{author}{%
  \bibinfo {author} {\bibfnamefont{A.~J.}\ \bibnamefont{Sadlej}}\ and\ \bibinfo
  {author} {\bibfnamefont{M.}~\bibnamefont{Urban}},\ }%
  \bibfield{journal}{%
  \bibinfo {journal} {J. Molec. Struct. THEOCHEM}\ }%
  \textbf{\bibinfo {volume} {234}},\ \bibinfo {pages} {147} (\bibinfo {year}
  {1991})%
  \bibAnnoteFile{NoStop}{sadlej1991}%
\bibitem{deiglmayr2008}%
  \BibitemOpen
  \bibfield{author}{%
  \bibinfo {author} {\bibfnamefont{J.}~\bibnamefont{Deiglmayr}}, \bibinfo
  {author} {\bibfnamefont{M.}~\bibnamefont{Aymar}}, \bibinfo {author}
  {\bibfnamefont{R.}~\bibnamefont{Wester}}, \bibinfo {author}
  {\bibfnamefont{M.}~\bibnamefont{Weidem\"uller}},\ and\ \bibinfo {author}
  {\bibfnamefont{O.}~\bibnamefont{Dulieu}},\ }%
  \bibfield{journal}{%
  \bibinfo {journal} {J. Chem. Phys.}\ }%
  \textbf{\bibinfo {volume} {129}},\ \bibinfo {pages} {064309} (\bibinfo {year}
  {2008})%
  \bibAnnoteFile{NoStop}{deiglmayr2008}%
\bibitem{sternheimer1969}%
  \BibitemOpen
  \bibfield{author}{%
  \bibinfo {author} {\bibfnamefont{R.~M.}\ \bibnamefont{Sternheimer}},\ }%
  \bibfield{journal}{%
  \bibinfo {journal} {Phys. Rev.}\ }%
  \textbf{\bibinfo {volume} {183}},\ \bibinfo {pages} {112} (\bibinfo {year}
  {1969})%
  \bibAnnoteFile{NoStop}{sternheimer1969}%
\bibitem{schwartz1974}%
  \BibitemOpen
  \bibfield{author}{%
  \bibinfo {author} {\bibfnamefont{H.~L.}\ \bibnamefont{Schwartz}}, \bibinfo
  {author} {\bibfnamefont{T.~M.}\ \bibnamefont{Miller}},\ and\ \bibinfo
  {author} {\bibfnamefont{B.}~\bibnamefont{Bederson}},\ }%
  \bibfield{journal}{%
  \bibinfo {journal} {Phys. Rev. A}\ }%
  \textbf{\bibinfo {volume} {10}},\ \bibinfo {pages} {1924} (\bibinfo {year}
  {1974})%
  \bibAnnoteFile{NoStop}{schwartz1974}%
\bibitem{derevianko2010}%
  \BibitemOpen
  \bibfield{author}{%
  \bibinfo {author} {\bibfnamefont{A.}~\bibnamefont{Derevianko}}, \bibinfo
  {author} {\bibfnamefont{S.~G.}\ \bibnamefont{Porsev}},\ and\ \bibinfo
  {author} {\bibfnamefont{J.~F.}\ \bibnamefont{Babb}},\ }%
  \bibfield{journal}{%
  \bibinfo {journal} {At. Data Nucl. Data Tables}\ }%
  \textbf{\bibinfo {volume} {96}},\ \bibinfo {pages} {323} (\bibinfo {year}
  {2010})%
  \bibAnnoteFile{NoStop}{derevianko2010}%
\bibitem{desclaux1981}%
  \BibitemOpen
  \bibfield{author}{%
  \bibinfo {author} {\bibfnamefont{J.-P.}\ \bibnamefont{Desclaux}}, \bibinfo
  {author} {\bibfnamefont{L.}~\bibnamefont{Laaksonen}},\ and\ \bibinfo {author}
  {\bibfnamefont{P.}~\bibnamefont{Pyykko}},\ }%
  \bibfield{journal}{%
  \bibinfo {journal} {J. Phys. B: At. Mol. Opt. Phys.}\ }%
  \textbf{\bibinfo {volume} {14}},\ \bibinfo {pages} {419} (\bibinfo {year}
  {1981})%
  \bibAnnoteFile{NoStop}{desclaux1981}%
\bibitem{fricke1986}%
  \BibitemOpen
  \bibfield{author}{%
  \bibinfo {author} {\bibfnamefont{B.}~\bibnamefont{Fricke}},\ }%
  \bibfield{journal}{%
  \bibinfo {journal} {J. Chem. Phys.}\ }%
  \textbf{\bibinfo {volume} {84}},\ \bibinfo {pages} {862} (\bibinfo {year}
  {1986})%
  \bibAnnoteFile{NoStop}{fricke1986}%
\bibitem{thorhallsson1968}%
  \BibitemOpen
  \bibfield{author}{%
  \bibinfo {author} {\bibfnamefont{J.}~\bibnamefont{Thorhallsson}}, \bibinfo
  {author} {\bibfnamefont{C.}~\bibnamefont{Fisk}},\ and\ \bibinfo {author}
  {\bibfnamefont{S.}~\bibnamefont{Fraga}},\ }%
  \bibfield{journal}{%
  \bibinfo {journal} {J. Chem. Phys.}\ }%
  \textbf{\bibinfo {volume} {49}},\ \bibinfo {pages} {1987} (\bibinfo {year}
  {1968})%
  \bibAnnoteFile{NoStop}{thorhallsson1968}%
\bibitem{stwalley2010}%
  \BibitemOpen
  \bibfield{author}{%
  \bibinfo {author} {\bibfnamefont{W.~C.}\ \bibnamefont{Stwalley}}, \bibinfo
  {author} {\bibfnamefont{J.}~\bibnamefont{Banerjee}}, \bibinfo {author}
  {\bibfnamefont{M.}~\bibnamefont{Bellos}}, \bibinfo {author}
  {\bibfnamefont{R.}~\bibnamefont{Carollo}}, \bibinfo {author}
  {\bibfnamefont{M.}~\bibnamefont{Recore}},\ and\ \bibinfo {author}
  {\bibfnamefont{M.}~\bibnamefont{Mastroianni}},\ }%
  \bibfield{journal}{%
  \bibinfo {journal} {J. Phys. Chem. A}\ }%
  \textbf{\bibinfo {volume} {114}},\ \bibinfo {pages} {81} (\bibinfo {year}
  {2010})%
  \bibAnnoteFile{NoStop}{stwalley2010}%
\bibitem{krych2010}%
  \BibitemOpen
  \bibfield{author}{%
  \bibinfo {author} {\bibfnamefont{M.}~\bibnamefont{Krych}}, \bibinfo {author}
  {\bibfnamefont{W.}~\bibnamefont{Skomorowski}}, \bibinfo {author}
  {\bibfnamefont{F.}~\bibnamefont{Pawlowski}}, \bibinfo {author}
  {\bibfnamefont{R.}~\bibnamefont{Moszynski}},\ and\ \bibinfo {author}
  {\bibfnamefont{Z.}~\bibnamefont{Idziaszek}},\ }%
  \bibfield{journal}{%
  \bibinfo {journal} {arXiv:1008.0840v1 [physics.atom-ph]}}%
   (\bibinfo {year} {2010})%
  \bibAnnoteFile{NoStop}{krych2010}%
\bibitem{aymar2009}%
  \BibitemOpen
  \bibfield{author}{%
  \bibinfo {author} {\bibfnamefont{M.}~\bibnamefont{Aymar}}, \bibinfo {author}
  {\bibfnamefont{J.}~\bibnamefont{Deiglmayr}},\ and\ \bibinfo {author}
  {\bibfnamefont{O.}~\bibnamefont{Dulieu}},\ }%
  \bibfield{journal}{%
  \bibinfo {journal} {Can. J. Phys}\ }%
  \textbf{\bibinfo {volume} {87}},\ \bibinfo {pages} {5443} (\bibinfo {year}
  {2009})%
  \bibAnnoteFile{NoStop}{aymar2009}%
\bibitem{aymar2009a}%
  \BibitemOpen
  \bibfield{author}{%
  \bibinfo {author} {\bibfnamefont{M.}~\bibnamefont{Aymar}}, \bibinfo {author}
  {\bibfnamefont{R.~G.}\ \bibnamefont{nd~M.~Sahlaoui}},\ and\ \bibinfo {author}
  {\bibfnamefont{O.}~\bibnamefont{Dulieu}},\ }%
  \bibfield{journal}{%
  \bibinfo {journal} {J. Phys. B}\ }%
  \textbf{\bibinfo {volume} {42}},\ \bibinfo {pages} {154025} (\bibinfo {year}
  {2009})%
  \bibAnnoteFile{NoStop}{aymar2009a}%
\bibitem{kolos1967}%
  \BibitemOpen
  \bibfield{author}{%
  \bibinfo {author} {\bibfnamefont{W.}~\bibnamefont{Ko\l{os}}}\ and\ \bibinfo
  {author} {\bibfnamefont{L.}~\bibnamefont{Wolniewicz}},\ }%
  \bibfield{journal}{%
  \bibinfo {journal} {J. Chem. Phys.}\ }%
  \textbf{\bibinfo {volume} {46}},\ \bibinfo {pages} {1426} (\bibinfo {year}
  {1967})%
  \bibAnnoteFile{NoStop}{kolos1967}%
\bibitem{moore1958}%
  \BibitemOpen
  \bibfield{author}{%
  \bibinfo {author} {\bibfnamefont{C.}~\bibnamefont{Moore}},\ }%
  \emph{\bibinfo {title} {Atomic energy levels, vol. 3}}\ (\bibinfo {publisher}
  {US Government printing office},\ \bibinfo {address} {Washington},\ \bibinfo
  {year} {1958})%
  \bibAnnoteFile{NoStop}{moore1958}%
\bibitem{hudson2009}%
  \BibitemOpen
  \bibfield{author}{%
  \bibinfo {author} {\bibfnamefont{E.~R.}\ \bibnamefont{Hudson}},\ }%
  \bibfield{journal}{%
  \bibinfo {journal} {Phys. Rev. A}\ }%
  \textbf{\bibinfo {volume} {79}},\ \bibinfo {pages} {032716} (\bibinfo {year}
  {2009})%
  \bibAnnoteFile{NoStop}{hudson2009}%
\end{thebibliography}

%

\end{document}